\documentclass[reqno, 12pt, psamsfonts]{amsart}
\usepackage[utf8]{inputenc}
\usepackage{datetime}
\usepackage{graphicx}
\usepackage{multicol}
\usepackage{pstricks}
\usepackage{amsmath}

\allowdisplaybreaks

\newcommand{\sI}{{\mathcal I}}
\newcommand{\sF}{{\mathcal F}}
\newcommand{\sC}{{\mathcal C}}
\newcommand{\sS}{{\mathcal S}}
\newcommand{\sA}{{\mathcal A}}
\newcommand{\OPT}{\mathrm{OPT}}

\newcommand{\OR}{\mathrm{OR}}

\newcommand{\up}[1]{\raisebox{1.6ex}[0pt]{#1}}

\newcommand{\CISC}{5}
\newcommand{\CISM}{360}
\newcommand{\CISD}{0.325309}

\newcommand{\CISFactorBounded}{2.3536}
\newcommand{\CISFactorUnbounded}{2.3105}

\newcommand{\CISAA}{551}
\newcommand{\CISAB}{600}
\newcommand{\CISAC}{43.1}
\newcommand{\CISAD}{18.48}

\newcommand{\CISORsmall}{0.729189}

\newcommand{\CISseqCtwo}{72}
\newcommand{\CISWoneCtwo}{35}
\newcommand{\CISWtwoCtwo}{37}

\newcommand{\CISseqCfour}{34}
\newcommand{\CISWoneCfour}{8}
\newcommand{\CISWtwoCfour}{9}

\newcommand{\CIIRTC}{3}
\newcommand{\CIIRTM}{93}
\newcommand{\CIIRTD}{0.229661}

\newcommand{\CIIRTFactorBounded}{2.5490}
\newcommand{\CIIRTFactorUnbounded}{2.5094}
\newcommand{\CIIRTLowerBound}{2.1193}

\newcommand{\CIIRTAA}{55}
\newcommand{\CIIRTAB}{144}
\newcommand{\CIIRTAC}{25.86}
\newcommand{\CIIRTAD}{32.17}

\newcommand{\CIIRTORsmall}{0.246036}

\newcommand{\CIIRTseqCtwo}{63}
\newcommand{\CIIRTWoneCtwo}{29}
\newcommand{\CIIRTWtwoCtwo}{34}

\newcommand{\SICC}{3}
\newcommand{\SICM}{300}
\newcommand{\SICD}{0.33213}

\newcommand{\SICFactorBounded}{3.5316}
\newcommand{\SICFactorUnbounded}{3.5146}
\newcommand{\SICLowerBound}{2.7707}

\newcommand{\SICAA}{649}
\newcommand{\SICAB}{702}
\newcommand{\SICAC}{40.98}
\newcommand{\SICAD}{61.26}
\newcommand{\SICAE}{2443.77}

\newcommand{\SICORsmall}{0.585590}

\newcommand{\SICseqStwo}{28}
\newcommand{\SICWoneStwo}{14}
\newcommand{\SICWtwoStwo}{15}

\newcommand{\SICseqSfour}{18}
\newcommand{\SICWoneSfour}{4}
\newcommand{\SICWtwoSfour}{5}

\usepackage{setspace}
\usepackage{paralist}
\usepackage{thmtools,thm-restate}
\usepackage[noend]{algpseudocode}
\usepackage{gensymb}
\usepackage[colorlinks=false]{hyperref}

\newtheorem{theorem}             {Theorem}[section]

\allowdisplaybreaks

\author[C. N. Lintzmayer]{Carla Negri Lintzmayer}
\author[F. K. Miyazawa]{Flávio Keidi Miyazawa}
\author[E. C. Xavier]{Eduardo Candido Xavier}

\address{Instituto de Computação, Universidade Estadual de Campinas, Campinas,
Brazil}
\email{\{carlanl,fkm,eduardo\}@ic.unicamp.br}

\thanks{This work was supported by São Paulo Research Foundation (grants
2016/14132-4, 2015/11937-9) and National Counsel of Technological and Scientific
Development (grants 306358/2014-0, 311499/2014-7, and 425340/2016-3).}

\begin{document}
\onehalfspacing
\date{\today, \currenttime}

\title{Online Circle and Sphere Packing}

\maketitle

\begin{abstract}
    In this paper we consider the Online Bin Packing Problem in three variants:
    Circles in Squares, Circles in Isosceles Right Triangles, and Spheres in
    Cubes.
    The two first ones receive an online sequence of circles (items) of
    different radii while the third one receive an online sequence of spheres
    (items) of different radii, and they want to pack the items into the minimum
    number of unit squares, isosceles right triangles of leg length one, and
    unit cubes, respectively.
    For Online Circle Packing in Squares, we improve the previous best-known
    competitive ratio for the bounded space version, when at most a constant
    number of bins can be open at any given time, from $2.439$ to
    $\CISFactorBounded$.
    For Online Circle Packing in Isosceles Right Triangles and Online Sphere
    Packing in Cubes we show bounded space algorithms of asymptotic competitive
    ratios $\CIIRTFactorBounded$ and $\SICFactorBounded$, respectively, as well
    as lower bounds of $\CIIRTLowerBound$ and $\SICLowerBound$ on the
    competitive ratio of any online bounded space algorithm for these two
    problems.
    We also considered the online unbounded space variant of these three
    problems which admits a small reorganization of the items inside the bin after
    their packing, and we present algorithms of competitive ratios
    $\CISFactorUnbounded$, $\CIIRTFactorUnbounded$, and $\SICFactorUnbounded$
    for Circles in Squares, Circles in Isosceles Right Triangles, and Spheres in
    Cubes, respectively.
\end{abstract}


\section{Introduction}
\label{sec:introduction}

In two or three-dimensional (offline) bin packing problems, one receives a list
of items and wants to pack them into the minimum number of bins so that two
items do not overlap and each item must be totally contained in the bin.
In this paper, we are interested in problems where the items can be circles or
spheres.
Thus, packing a circle $i$ of radius $r_i$ means finding coordinates $x_i$ and
$y_i$ to its center such that they respect the boundaries of the bin and, for
every other circle $j$ of radius $r_j$ and coordinates $x_j$ and $y_j$,
$(x_i-x_j)^2 + (y_i-y_j)^2 \geq (r_i+r_j)^2$.
For spheres this is similar, but we also have coordinates $z_i$.

As applications for these problems where items can be circles or spheres, we can
mention origami design~\cite{2010-demaine-etal}, crystallography,
error-correcting codes, coverage of a geographical area with cell transmitters,
storage of cylindrical barrels, and packaging bottles or
cans~\cite{2007-szabo-etal}.

In the online variant, one item arrives at a time and must be packed as soon as
it arrives, without knowledge of further items.
Also, after packing an item, it cannot be moved to another bin.
At any moment we consider that a bin is open or closed.
We can only pack items into open bins, and once a bin is closed, it cannot be
opened again.
With this, we can classify algorithms as having \textit{bounded space}, when the
number of open bins can be bounded by a constant, or \textit{unbounded space},
when there is no guarantee on the number of open bins.
In this paper, we do not allow items packed in a same bin to be reorganized,
except for the unbounded space algorithms, which may do this on a constant
number of items.

We consider the Online Circle Packing in Squares problem, the Online Circle
Packing in Isosceles Right Triangles problem, and the Online Sphere Packing in
Cubes problem.
The two first ones receive an online sequence of circles of different radii and
the objective is to pack them into the minimum number of unit squares and
isosceles right triangles of leg length one, respectively.
Similarly, the third one receives an online sequence of spheres of different
radii and the objective is to pack them into the minimum number of unit cubes.

Demaine \textit{et al.}~\cite{2010-demaine-etal} showed that it is NP-hard to
decide whether a given set of circles can be packed into a rectangle, an
equilateral triangle, or a unit square.
Still, we are interested in designing algorithms that can give some guarantee on
the number of bins used.
We say that an online algorithm $\sA$ has \textit{competitive ratio} $\rho$ if
$\sA(I) \leq \rho\OPT(I)$ for every instance $I$, where $\sA(I)$ is the number
of bins used by the algorithm and $\OPT(I)$ is the number of bins of an optimal
offline solution.
If $\sA(I) \leq \rho\OPT(I) + c$ for a constant $c$, then we say the algorithm
has \textit{asymptotic competitive ratio} $\rho$.

When the items of the packing problem are not circles or spheres, there are a
number of results in the literature.
Friedman~\cite{2009-friedman}, for instance, considers packing squares into
squares, while Kamali \textit{et al.}~\cite{2015-kamali-etal} consider packing
equilateral triangles into squares.
More common results, though, consider packing rectangles into unit squares.
We highlight the best-known algorithms for this problem, for which we can yet
consider the case when orthogonal rotations of the items are allowed or not.
For the offline version with or without rotations, the best asymptotic
approximation has factor $\ln(1.5)+1 \approx 1.405$~\cite{2014-bansal-khan}.
For the online version, the best asymptotic competitive ratio is $2.5467$ for
the bounded space case~\cite{2010-epstein} and $2.25$ for the unbounded space
case~\cite{2006-epstein-vanstee}.
For three dimensions, there exists an asymptotic polynomial time approximation
scheme for packing cubes into unit cubes~\cite{2006-bansal-etal} and a
competitive algorithm of asymptotic ratio~$2.6161$ for the online
version~\cite{2010-han-etal}. 
A recent survey by Christensen \textit{et al.}~\cite{2017-christensen-etal}
presents approximation and online algorithms for some variations of bin packing
in two or three dimensions.
A survey by Lodi \textit{et al.}~\cite{2013-lodi-etal} also considers
two-dimensional variations, contemplating heuristics and exact algorithms.

To the best of our knowledge, competitive algorithms for the Online Circle
Packing in Squares were considered only by Hokama~\textit{et
al.}~\cite{2016-hokama-etal}, who showed a lower bound of $2.292$ on the
competitive ratio of any bounded space algorithm and gave one such algorithm
with asymptotic competitive ratio~$2.439$.
We did not find results within this approach for Online Circle Packing in
Isosceles Right Triangles and Online Sphere Packing in Cubes.

For the offline Circle Packing in Squares, Miyazawa \textit{et
al.}~\cite{2016-miyazawa-etal} showed an asymptotic polynomial time
approximation scheme when one can augment one direction of the unit bin by a
small constant.
There are also several results that involve packing circles of same radii into
squares~\cite{2007-szabo-etal} or packing spheres of same radii into
cubes~\cite{2015-tatarevic}.
We also refer the reader to \cite{2009-hifi-mhallah} for a review on circle and
sphere packing related results.

\textit{Our contributions.} We show a general idea of bounded space and
unbounded space algorithms for packing circles/spheres which works for the three
problems that we are considering.
The bounded space algorithms are in part based on the one given by
Hokama~\textit{et al.}~\cite{2016-hokama-etal}, but they have a simpler
analysis.
Furthermore, we were able to improve the occupation ratio related to the class
of small circles, we found a simpler way of subdividing a bin, and we used
another method to find the numerical results.
With that, our bounded space algorithm for Online Circle Packing in Squares has
asymptotic competitive ratio $\CISFactorBounded$, for Online Circle Packing in
Isosceles Right Triangles it has asymptotic competitive ratio
$\CIIRTFactorBounded$, and for Online Sphere Packing in Cubes it has asymptotic
competitive ratio $\SICFactorBounded$.
The unbounded space algorithms are a modification of the bounded space ones
using an idea presented by Epstein~\cite{2010-epstein}.
Our unbounded space algorithm for Online Circle Packing in Squares has
asymptotic competitive ratio $\CISFactorUnbounded$, for Online Circle Packing in
Isosceles Right Triangles it has asymptotic competitive ratio
$\CIIRTFactorUnbounded$, and for Online Sphere Packing in Cubes it has
asymptotic competitive ratio $\SICFactorUnbounded$.
These algorithms allow a small reorganization of the items after their packing
inside the bin (they do not allow items to be exchanged between bins).
Lastly, we give lower bounds of $\CIIRTLowerBound$ and $\SICLowerBound$ on the
competitive ratio of any bounded space algorithm for Online Circle Packing in
Isosceles Right Triangles and Online Sphere Packing in Cubes, respectively.
Our results are summarized in Table~\ref{tab:results}.
We highlight that the asymptotic competitive ratio given by our bounded space
algorithm for Online Circle Packing in Squares is an improvement on a previous
result while all other results we give are new.

\begin{table}[h]
\renewcommand{\arraystretch}{1.5}
  \caption{Summary of the results presented in this paper.}
  \label{tab:results}
  \scriptsize
  \centering
  \begin{tabular}{|c|c|c|c|}
    \hline
    & \multicolumn{3}{c|}{\textbf{Competitive ratio}} \\ \cline{2-4}
    \up{\textbf{Problem}} & \textbf{Lower bound} & \textbf{Bounded space} &
    \textbf{Unbounded space} \\ \hline
    Circle Packing in Squares & $2.292$~\cite{2016-hokama-etal} &
    \CISFactorBounded~(Thm. \ref{thm:cis:bounded_ratio}) &
    \CISFactorUnbounded~(Thm. \ref{thm:cis:unbounded_ratio}) \\ \hline
    Circle Packing in Isosceles Right Triangle & \CIIRTLowerBound~(Thm.
    \ref{thm:ciirt:lower_bound}) & \CIIRTFactorBounded~(Thm.
    \ref{thm:ciirt:bounded_ratio}) & \CIIRTFactorUnbounded~(Thm.
    \ref{thm:ciirt:unbounded_ratio})
    \\ \hline
    Sphere Packing in Cubes & \SICLowerBound~(Thm. \ref{thm:sic:lower_bound}) &
    \SICFactorBounded~(Thm. \ref{thm:sic:bounded_ratio}) &
    \SICFactorUnbounded~(Thm. \ref{thm:sic:unbounded_ratio}) \\ \hline
  \end{tabular}
\end{table}

Section~\ref{sec:general} describes the general idea of our algorithms.
Sections~\ref{sec:cis}, \ref{sec:ciirt}, and \ref{sec:sic} specify the
algorithms for each of the three problems.

\section{Preliminaries and general ideas}
\label{sec:general}

In this section, we use the word ``\textit{item}'' to denote a circle or a
sphere, according to the problem.
A ``\textit{unit bin}'' means a unit square, an isosceles right triangle of leg
length one, or a unit cube, also according to the problem.

The general idea of the bounded and unbounded space algorithms that we will
present is the following.
We classify items into large ones and small ones and pack them separately so
that we can differentiate \textit{Lbins}, which are bins that pack large items,
from \textit{Sbins}, which are bins that pack small items.
Nevertheless, Lbins and Sbins are unit bins.
We will further classify the items and bins in types so that large items of type
$i$ are packed into Lbins of type $i$ and small items of type $i$ are packed
into Sbins of type $i$, for some integer $i$.
After explaining the algorithms, we will analyse their asymptotic competitive
ratios using the \textit{weighting method}~\cite{2010-epstein}, which will be
explained later.
For that, the concept of \textit{occupation ratio} of a closed bin, which is the
minimum total area/volume occupied by the items packed in such bin, will be
notably used.

\begin{figure}[h]
    \centering
    \resizebox{\textwidth}{!}{\input{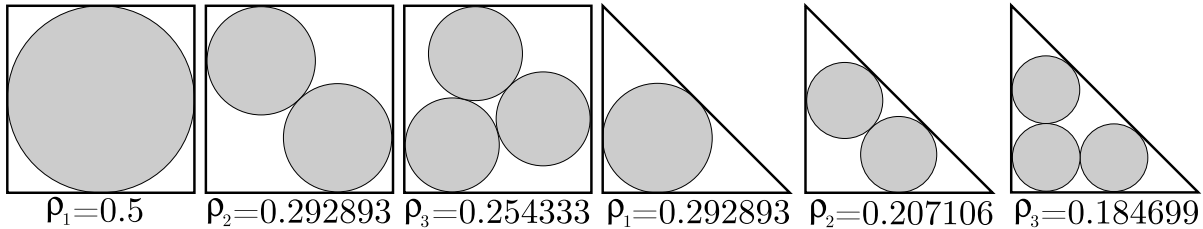}}
    \caption{Values of $\rho_1$, $\rho_2$, and $\rho_3$ for unit squares and
    isosceles right triangles of leg length one.}
    \label{fig:rhos}
\end{figure}

For an integer $M$ given as a parameter, an item of radius $r$ is \textit{large}
if $r > 2/M$ and it is \textit{small} if $r \leq 2/M$.
For a positive integer $i \geq 1$, let $\rho_i*$ be the largest value of radius
such that $i$ items of radius $\rho_i*$ can be packed into a unit bin.  Let
$\rho_i$ be $\rho_i*$ if the latter is known or the best-known lower bound for
$\rho_i*$~\cite{packomania,2007-szabo-etal,2015-tatarevic}, otherwise.
See Figure~\ref{fig:rhos} for some examples.

Let $K$ be an integer such that $\rho_{K+1} < 2/M \leq \rho_K$.
We classify a large item of radius $r$ as type $i$, for $1 \leq i < K$, if
$\rho_{i+1} < r \leq \rho_i$ and as type $K$ if $2/M < r \leq \rho_K$.
We will use $I_i$ to denote a large item of type $i$.

Let $C$ be an integer given as a parameter.
Given a small item of radius $r$, we find the largest integer $p$ such that $C^p
r \leq 2/M$.
Then we classify such item as type $i$ and subtype $p$ if $2/(i+1) < C^p r \leq
2/i$, where $M \leq i < CM$.
For simplicity, we will say that the item is of type $(i,p)$.

The hexagonal packing is the densest one for circles of equal radii and the
hexagonal close packing is the densest one for spheres of equal radii.
The first one, of density $\pi/\sqrt{12}$, can be seen as if we tessellate the
two dimensional space with hexagons and place a circle in each one of them.
The second one, of density $\pi/\sqrt{18}$, can be seen as if we tessellate the
three dimensional space with rhombic dodecahedra and place a sphere in each one
of them.
We will use these tessellations to pack small items in two or three dimensions,
accordingly.
For that, note that a regular hexagon of side length $2r/\sqrt{3}$ has an
inscribed circle of radius at most $r$ and a rhombic dodecahedron of side length
$3r/\sqrt{6}$ has an inscribed sphere of radius at most~$r$.

\begin{figure}[h]
    \centering
    \resizebox{\textwidth}{!}{\input{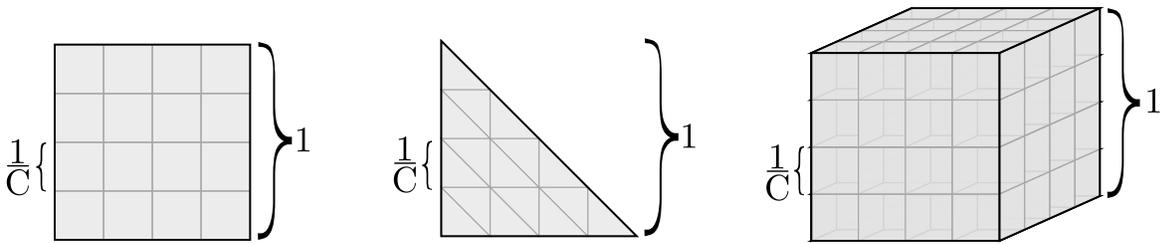}}
    \caption{Division of an Sbin of type $i$ into q-bins$(i,0)$ for $C = 4$.}
    \label{fig:initialize_sbin}
\end{figure}

Both bounded and unbounded space algorithms pack small items in the same way, so
we start by describing this part of them.
We define a \textit{q-bin(i,p)} as a polytope that has the same form of the bin,
but has side length $1/C^{p+1}$.
For each $i$, with $M \leq i < CM$, at most one Sbin of type $i$ is kept open at
a time to pack small circles of type $(i,p)$, for any $p \geq 0$.
When an Sbin of type $i$ is opened, it is divided into $C^d$ q-bins$(i,0)$,
where $d$ is the dimension of the space.
See Figure~\ref{fig:initialize_sbin}.

\begin{figure}[h]
    \centering
    \resizebox{\textwidth}{!}{\input{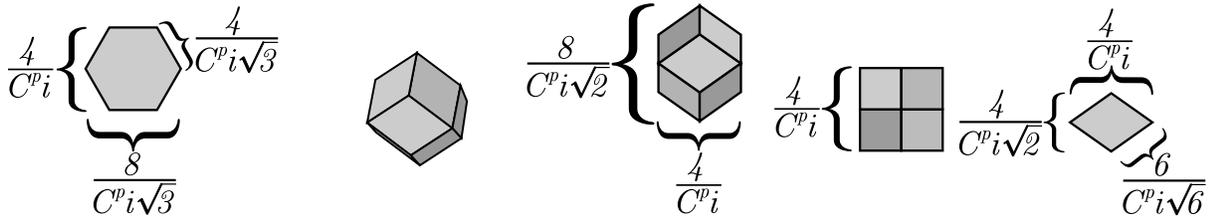}}
    \caption{Measurements of the cells used to tessellate a q-bin$(i,p)$.}
    \label{fig:cells}
\end{figure}

Let $\ell = 4/(C^pi\sqrt{3})$ if the items are circles, let $\ell =
6/(C^pi\sqrt{6})$ if they are spheres, and let $d$ be the dimension of the
space.
We will use the word ``cell'' to generally denote a hexagon or a rhombic
dodecahedron of a tessellation.
A q-bin$(i,p)$ is either subdivided into $C^d$ q-bins$(i,p+1)$ or it is
tessellated with cells of side $\ell$ to pack small items of type $(i,p)$.
Note that we must have $1/C^{p+1} \geq 2/(iC^p)$ for this to work, but since $i
< CM$, it suffices to choose a $M > 2$.
Also, note that $\ell$ is chosen precisely so that the cell can pack an item of
type $(i,p)$.
Figure~\ref{fig:cells} shows the measurements of the cells.

For two dimensions, the tessellation of the q-bin$(i,p)$ is done by placing a
hexagon on the left bottom corner of the bin with two of its sides parallel to
the bottom of the bin.
For three dimensions, the tessellation is done by placing a rhombic dodecahedron
in the left bottom corner of the bin with four of its faces, each one sharing
two nonadjacent vertices with two of the other, parallel to the sides of the
bin.
See Figure~\ref{fig:qbin(i,p)}.
Note that hexagons and rhombic dodecahedra are space-filling polytopes.

\begin{figure}[h]
    \centering
    \resizebox{\textwidth}{!}{\input{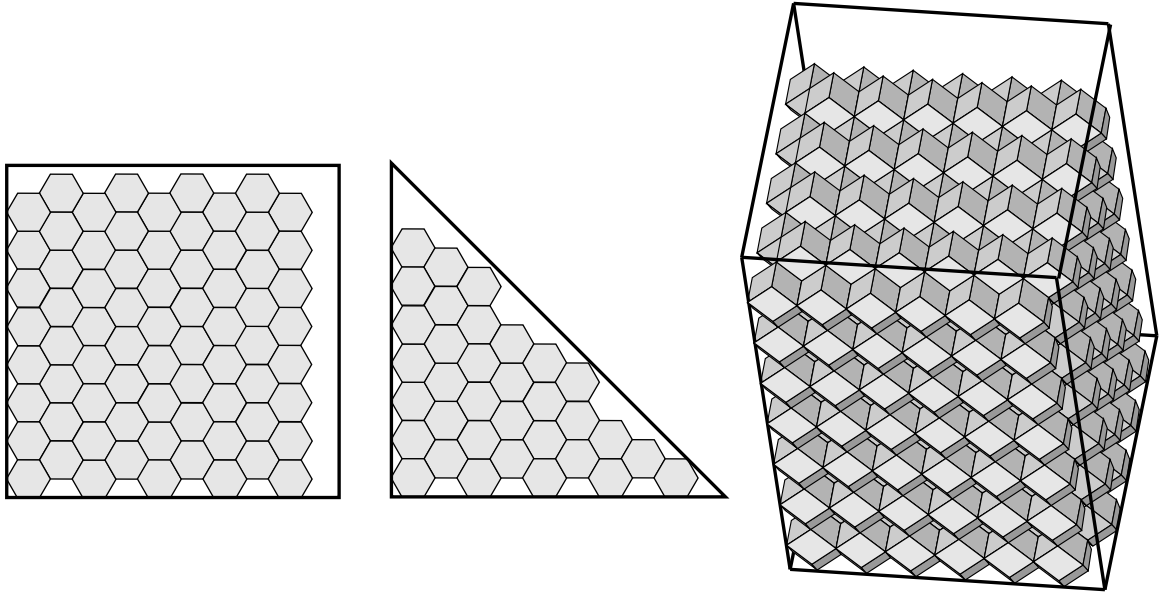}}
    \caption{Tesselation of a q-bin$(i,p)$.}
    \label{fig:qbin(i,p)}
\end{figure}

When a small item of type $(i,p)$ arrives, the algorithms will simply try to
pack it into a cell of a q-bin$(i,p)$. 
If there is no such cell, they will subdivide a larger q-bin$(i,p')$, i.e., find
an empty q-bin$(i,p')$ with the largest $p'$ such that $p' < p$ and subdivide
it, until a q-bin$(i,p)$ is found.
This is formally described below:
\begin{algorithmic}
\If {there is no tessellated q-bin$(i,p)$ with empty cells}
    \If {there is no empty q-bin$(i,p)$ or empty q-bin$(i,p')$ with $p'<p$}
        \State close the current Sbin of type $i$ (if any) and open a new one;
    \EndIf
    \While {there is no empty q-bin$(i,p)$}
        \State let $p'$ be the largest integer $< p$ such that there is an empty
        q-bin$(i,p')$;
        \State subdivide one empty q-bin$(i,p')$ in $C^d$ q-bins$(i,p'+1)$;
    \EndWhile
    \State tessellate an empty q-bin$(i,p)$;
\EndIf
\State pack the item in one empty cell of a tessellated q-bin$(i,p)$.
\end{algorithmic}

\begin{figure}[h]
    \centering
    \resizebox{\textwidth}{!}{\input{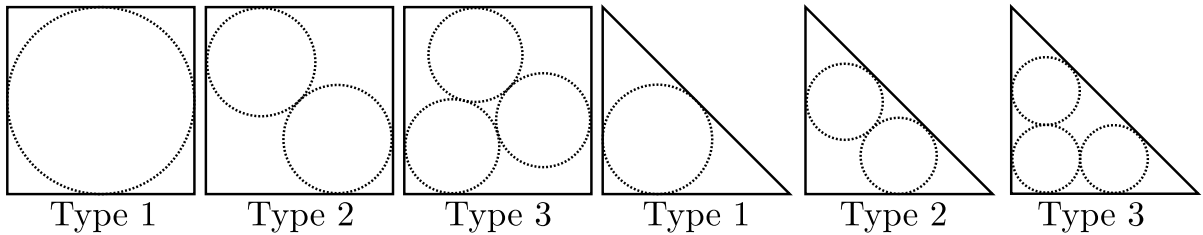}}
    \caption{Three first types of Lbins for unit squares and isosceles right
    triangles of leg length one.}
    \label{fig:cbins}
\end{figure}

Now we consider packing large circles in the bounded space algorithms.
For each $i$, with $1 \leq i \leq K$, at most one Lbin of type $i$ is kept open
at a time to pack large items of type $i$.
When an Lbin of type $i$ is opened, it is divided into $i$ \textit{c-bins},
which are circles/spheres of radii $\rho_i$.
See Figure~\ref{fig:cbins}.

When an $I_i$ arrives, it is either packed into an empty c-bin or the current
Lbin of type $i$ (if any) is closed and a new one is opened.
Note that at most $K$ Lbins are kept open by the algorithm at any given time.
Together with the $(C-1)M$ Sbins, we can see that this algorithm has bounded
space. 

Packing large circles in the unbounded space algorithms follows an idea
given by Epstein~\cite{2010-epstein}, which uses \textit{waiting bins}.
A waiting bin packs an $I_1$ with other items, instead of packing it alone, if
their radii are related with a given parameter $D$.
All large items which are not packed in waiting bins are packed as in the
bounded space algorithms.
As we mentioned, the unbounded space algorithms allow some reorganization of the
items after their packing.
We highlight here that this happens only for items $I_1$ that are inside open
waiting bins.
Since the specific items and possible configurations of waiting bins depend on
the format of the bin, we will better describe such algorithms in each of the
following sections.
Also, we cannot guarantee how many waiting bins are kept open by the algorithms
at any given time, which is why they have unbounded space.

Each of the next sections analyses the asymptotic competitive ratios of the
algorithms considering Online Circle Packing in Squares, Online Circle Packing
in Isosceles Right Triangles, and Online Sphere Packing into Cubes.
We specify parameters $M$, $C$, and $D$ for each problem as well as give more
details on the unbounded space algorithms.
We also present lower bounds on the competitive ratios of any bounded space
algorithm for Online Circle Packing in Isosceles Right Triangles and Online
Sphere Packing into Cubes.

As we mentioned, the analysis of the asymptotic competitive ratio for the
bounded space algorithms uses the weighting method~\cite{2010-epstein}, in which
we first have to show a weighting function $w$ over the items such that the sum
of weights of items in any bin generated by the algorithms is at least 1, except
for a constant number of bins.
So, if $B$ is the set of items in a closed bin, $w(B) = \sum_{i \in B} w(i) \geq
1$. 
Afterwards, we find every possible feasible configuration, i.e., sets of items
that can be packed into a bin, calculate their sum of weights, and find the
supremum $\beta$ of such sum of weights.
As a result, the asymptotic competitive ratio of the algorithm is bounded from
above by $\beta$ because of the following. 
If $S$ is the total number of bins used by the algorithm, $X$ is the number of
closed bins at the end, and $O$ is the maximum number of open bins at the
end, then $S - O = X \leq w(\sI)$, where $\sI$ is the set of items of the input.
If $\OPT(\sI)$ is the number of bins used by an optimal offline solution, then
$w(\sI)/\beta \leq \OPT(\sI)$, since every feasible configuration has weight at
most $\beta$.
Thus, we have $S \leq w(\sI) + O \leq \beta\OPT(\sI)+O$.

The analysis of the asymptotic competitive ratio for the unbounded space
algorithms is made using a generalized form of the weighting
method~\cite{2010-epstein}, in which we have to show weighting functions $w_1$
and $w_2$ such that the sum of weights of items in any bin generated by the
algorithms is at least 1 on average for at least one of the functions, except
perhaps for a constant number of bins.
Again, the supremum $\beta$ of the sums of weights of feasible configurations,
considering both weight functions, is an upper bound for the asymptotic
competitive ratio.

Note that both analysis depend on finding the supremum of the sum of weights of
feasible configurations.
However, it is not reasonable to list every possible such set, so we use the
following fact.
Suppose that a bin of maximum weight has some known-items $\sI' \subseteq \sI$
whose sum of weights is $W$ and whose sum of areas/volumes is $V$, and let $\OR$
be a lower bound on the occupation ratio of the unknown items of the bin, i.e.,
items in $\sI \backslash \sI'$.
The weighting functions used in the next sections are such that the weight given
to any item $i$ not in $\sI'$ of area/volume $v(i)$ is $w(i) \leq v(i)/\OR$.  
Thus, the total sum of weights in such bin is at most $W + (S-V)/\OR$, where $S$
is the area/volume of the bin.

This means that we can concentrate on finding a few configurations and the lower
bound on the occupation ratio of items that are not on them, in order to find an
upper bound on the asymptotic competitive ratio of our algorithms.
For that, we use a two-phase program.
The first phase uses constraint programming to test if a given set of items is a
feasible or an infeasible configuration.
Given a set of items $\sI$ $=$ $\{i_1,$ $i_2,$ $\ldots,$ $i_n\}$, let $\sI'$ $=$
$\{i_1',$ $i_2',$ $\ldots,$ $i_n'\}$ be a set of items such that $r(i_j')$ $=$
$\mu r(i_j)$ for some $0 < \mu < 1$, where $r(i)$ is the radius of item $i$.  We
show in Theorems~\ref{thm:sic:feasible_infeasible},
\ref{thm:cis:feasible_infeasible}, and \ref{thm:cirrt:feasible_infeasible} that
the packing of $\sI'$ and $\sI$ are related for each of the three problems we
are considering.

\begin{theorem}
    \label{thm:sic:feasible_infeasible}
    Consider the Online Sphere Packing in Cubes problem.
    Let $\sS$ $=$ $\{s_1,$ $s_2,$ $\ldots,$ $s_n\}$ be a given set of spheres
    and $\sS'$ $=$ $\{s_1',$ $s_2',$ $\ldots,$ $s_n'\}$ be a set of spheres such
    that $r(s_i')$ $=$ $\mu r(s_i)$ for some $0 < \mu < 1$.
    Let $r_{min}$ $=$ $\min\{r(s_i) \colon s_i \in \sS\}$ and $\delta$ $\leq$
    $\frac{(1-\mu^2)r_{min}}{3}$.
    If $\sS'$ cannot be packed in a bin using a regular grid of granularity
    $\delta$, then $\sS$ cannot be packed in a bin with general positions.
\end{theorem}
\begin{proof}
    We will prove the contrapositive.
    Suppose $\sS$ can be packed in general positions of the bin and let
    $((x_1,y_1,z_1),$ $(x_2,y_2,z_2),$ $\ldots,$ $(x_n,y_n,z_n))$ be one of its
    packings.
    Let $((x_1',y_1',z_1'),$ $(x_2',y_2',z_2'),$ $\ldots,$ $(x_n',y_n',z_n'))$
    be a rounding of the previous packing to the nearest coordinates of the
    regular grid of granularity $\delta$.
    We will show that $((x_1',y_1',z_1'),$ $(x_2',y_2',z_2'),$ $\ldots,$
    $(x_n',y_n',z_n'))$ is a packing of $\sS'$.
    Note that, for every pair $s_i$ and $s_j$ of spheres,
    $|x_i' - x_j'| \geq |x_i - x_j| - \delta$,~ 
    $|y_i' - y_j'| \geq |y_i - y_j| - \delta$, and
    $|z_i' - z_j'| \geq |z_i - z_j| - \delta$, 
    because every coordinate has changed by at most $\delta/2$.

    Let $d$ be the distance between two spheres $s_i$ and $s_j$.
    Thus, $d^2 = (x_i - x_j)^2 + (y_i - y_j)^2 + (z_i - z_j)^2$,~ $d \geq r(s_i)
    + r(s_j) \geq 2r_{min}$,~ $|x_i - x_j| \leq d$,~ $|y_i - y_j| \leq d$, and
    $|z_i - z_j| \leq d$.

    There are four main cases.
    First, consider that $|x_i - x_j| \geq \delta$, $|y_i - y_j| \geq \delta$,
    and $|z_i - z_j| \geq \delta$. We have
    \setlength{\abovedisplayskip}{6pt}
    \setlength{\belowdisplayskip}{\abovedisplayskip}
    \setlength{\abovedisplayshortskip}{0pt}
    \setlength{\belowdisplayshortskip}{3pt}
    \begin{align*}
       |x_i' & - x_j'|^2 + |y_i' - y_j'|^2 + |z_i' - z_j'|^2 
            \geq (|x_i - x_j| - \delta)^2 + (|y_i - y_j| - \delta)^2 + (|z_i -
            z_j| - \delta)^2 \\
            &= d^2 - 2\delta(|x_i - x_j| + |y_i - y_j| + |z_i - z_j|) +
            3\delta^2 
            \geq d^2 - 6\delta d + 3\delta^2 \geq d^2 -6\delta d \\
            &\geq d^2 - 6\left(\frac{(1-\mu^2)r_{min}}{3}\right)d = d^2
            -2(1-\mu^2)r_{min}d \\
            &\geq d^2 - (1-\mu^2)d^2 = \mu^2d^2 \geq \mu^2(r(s_i) +
            r(s_j))^2 = (r(s_i') + r(s_j'))^2 \enspace,
    \end{align*}
    and thus $s_i'$ and $s_j'$ do not intersect in this case.

    Now consider, without loss of generality, that $|x_i - x_j| < \delta$, but
    $|y_i - y_j| \geq \delta$ and $|z_i - z_j| \geq \delta$.
    We have
    \setlength{\abovedisplayskip}{6pt}
    \setlength{\belowdisplayskip}{\abovedisplayskip}
    \setlength{\abovedisplayshortskip}{0pt}
    \setlength{\belowdisplayshortskip}{3pt}
    \begin{align*}
       |x_i' & - x_j'|^2 + |y_i' - y_j'|^2 + |z_i' - z_j'|^2 
            \geq (|y_i - y_j| - \delta)^2 + (|z_i - z_j| - \delta)^2 \\
            &= d^2 - |x_i - x_j|^2 - 2\delta(|y_i - y_j| + |z_i - z_j|) +
            2\delta^2 
            \geq d^2 - \delta^2 - 4\delta d + 2\delta^2 \geq d^2 - 4\delta d \\
            &\geq d^2 - 4\left(\frac{(1-\mu^2)r_{min}}{3}\right)d > d^2
            -2(1-\mu^2)r_{min}d \\
            &\geq d^2 - (1-\mu^2)d^2 = \mu^2d^2 \geq \mu^2(r(s_i) +
            r(s_j))^2 = (r(s_i') + r(s_j'))^2 \enspace,
    \end{align*}
    and thus $s_i'$ and $s_j'$ do not intersect in this case.

    For the third case, consider, without loss of generality, that $|x_i - x_j|
    < \delta$ and $|y_i - y_j| < \delta$, but $|z_i - z_j| \geq \delta$.
    We have
    \setlength{\abovedisplayskip}{6pt}
    \setlength{\belowdisplayskip}{\abovedisplayskip}
    \setlength{\abovedisplayshortskip}{0pt}
    \setlength{\belowdisplayshortskip}{3pt}
    \begin{align*}
       |x_i' & - x_j'|^2 + |y_i' - y_j'|^2 + |z_i' - z_j'|^2 
            \geq (|z_i - z_j| - \delta)^2 \\
            &= d^2 - |x_i - x_j|^2 - |y_i - y_j|^2 - 2 \delta |z_i - z_j| +
            \delta^2 
            \geq d^2 - 2\delta^2 - 2 \delta d + \delta^2 \geq d^2 - \delta^2 -
            \delta d \\
            &\geq d^2 - \frac{(1-\mu^2)^2r_{min}^2}{9} -
            2\left(\frac{(1-\mu^2)r_{min}}{3}\right)d \geq d^2 -
            \frac{(1-\mu^2)d^2}{36} - \frac{(1-\mu^2)d^2}{3} \\
            &= d^2 - \frac{13}{36}(1-\mu^2)d^2 > d^2 - (1-\mu^2)d^2 = \mu^2d^2
            \geq \mu^2(r(s_i) + r(s_j))^2 = (r(s_i') + r(s_j'))^2 \enspace,
    \end{align*}
    and thus $s_i'$ and $s_j'$ do not intersect in this case either.

    For the last case, consider that $|x_i - x_j| < \delta$, $|y_i - y_j| <
    \delta$, and $|z_i - z_j| < \delta$.
    We have
    \setlength{\abovedisplayskip}{6pt}
    \setlength{\belowdisplayskip}{\abovedisplayskip}
    \setlength{\abovedisplayshortskip}{0pt}
    \setlength{\belowdisplayshortskip}{3pt}
    \begin{multline*}
       |x_i - x_j|^2 + |y_i - y_j|^2 + |z_i - z_j|^2 < 3\delta^2 \leq
       3\left(\frac{(1-\mu^2)^2r_{min}^2}{9}\right) \leq
       \frac{(1-\mu^2)r_{min}^2}{3} \leq r_{min}^2 \enspace.
    \end{multline*}
    On the other hand,
    \setlength{\abovedisplayskip}{6pt}
    \setlength{\belowdisplayskip}{\abovedisplayskip}
    \setlength{\abovedisplayshortskip}{0pt}
    \setlength{\belowdisplayshortskip}{3pt}
    \begin{multline*}
       |x_i - x_j|^2 + |y_i - y_j|^2 + |z_i - z_j|^2 = d^2 \geq (r_i + r_j)^2
       \geq 4r_{min}^2 \enspace,
    \end{multline*}
    so we reach a contradiction, which means this case cannot happen.

    Now, it remains to show that every sphere $s_i'$ does not intercept the
    borders of the bin.
    Since $\frac{(1-\mu^2)r_{min}}{3} \leq 2(1-\mu)r_{min}$ for $\mu \leq 1$, we
    have that $\delta \leq 2(1-\mu)r_{min}$. 
    Thus,
{\footnotesize
  \setlength{\abovedisplayskip}{6pt}
  \setlength{\belowdisplayskip}{\abovedisplayskip}
  \setlength{\abovedisplayshortskip}{0pt}
  \setlength{\belowdisplayshortskip}{3pt}
    \begin{multline*}
    x_i' \leq x_i + \frac{\delta}{2} \leq 1 - r(s_i) + \frac{\delta}{2} \leq
            1-r(s_i) + (1-\mu)r_{min} \leq 1 - r(s_i) + (1-\mu)r(s_i) =
        1-\mu r(s_i) = 1-r(s_i')
    \end{multline*}
}%
    and
{\footnotesize
  \setlength{\abovedisplayskip}{6pt}
  \setlength{\belowdisplayskip}{\abovedisplayskip}
  \setlength{\abovedisplayshortskip}{0pt}
  \setlength{\belowdisplayshortskip}{3pt}
    \begin{align*}
    x_i' \geq x_i - \frac{\delta}{2} \geq r(s_i) - \frac{\delta}{2} \geq r(s_i)
            - (1\!-\!\mu)r_{min} \geq r(s_i) - (1\!-\!\mu)r(s_i) = \mu r(s_i) =
            r(s_i') \enspace.
    \end{align*}
}%
    The same is valid for coordinates $y$ and $z$. 

    Therefore, $((x_1',y_1',z_1'), (x_2',y_2',z_2'), \ldots, (x_n',y_n',z_n'))$
    is a packing of $\sS'$.
\end{proof}

\begin{theorem}
    \label{thm:cis:feasible_infeasible}
    Consider the Online Circle Packing in Squares problem.
    Let $\sC$ $=$ $\{c_1,$ $c_2,$ $\ldots,$ $c_n\}$ be a given set of circles
    and $\sC'$ $=$ $\{c_1',$ $c_2',$ $\ldots,$ $c_n'\}$ be a set of circles such
    that $r(c_i')$ $=$ $\mu r(c_i)$ for some $0 < \mu < 1$.
    Let $r_{min}$ $=$ $\min\{r(c_i) \colon c_i \in \sC\}$ and $\delta$ $\leq$
    $\frac{(1-\mu^2)r_{min}}{2}$.
    If $\sC'$ cannot be packed in a bin using a regular grid of granularity
    $\delta$, then $\sC$ cannot be packed in a bin with general positions.
\end{theorem}
\begin{proof}
    Similar to the proof of Theorem~\ref{thm:sic:feasible_infeasible}.
\end{proof}

\begin{theorem}
    \label{thm:cirrt:feasible_infeasible}
    Consider the Online Circle Packing in Isosceles Right Triangles problem.
    Let $\sC$ $=$ $\{c_1,$ $c_2,$ $\ldots,$ $c_n\}$ be a given set of circles
    and $\sC'$ $=$ $\{c_1',$ $c_2',$ $\ldots,$ $c_n'\}$ be a set of circles such
    that $r(c_i')$ $=$ $\mu r(c_i)$ for some $0 < \mu < 1$.
    Let $r_{min}$ $=$ $\min\{r(c_i) \colon c_i \in \sC\}$ and $\delta$ $\leq$
    $\frac{(1-\mu^2)r_{min}}{2}$.
    If $\sC'$ cannot be packed in a bin using a regular grid of granularity
    $\delta$, then $\sC$ cannot be packed in a bin with general positions.
\end{theorem}
\begin{proof}
    Similar to the proof of Theorem~\ref{thm:sic:feasible_infeasible}.
\end{proof}

Consider the notation given in the previous theorems.
In the first phase, to test for the feasibility of a packing, we rescale the
instance such that the size of the bin used in the test is $W = \lceil 1/\delta
\rceil$ and the radius of each item is also multiplied by $W$.
As we use an integer solver, we have to assure that all grid positions are
integers.
So, we take $r(i_j') = \lfloor r(i_j) \mu W\rfloor$ and we test if $\sI'$ can be
packed into a bin of side length $W$.
If not, then $\sI$ does not have a feasible packing into the bin of side length
$1$; otherwise there is no guarantee of infeasibility.
Also, we take $r(i_j') = \lceil r(i_j) W \rceil$ and we test if $\sI'$ can be
packed into the bin of side length $W$.
If it can, then $\sI$ can also be packed into the bin of side length $1$;
otherwise there is no guarantee of feasibility.

In the first phase we find a set $\sF$ of feasible configurations that contain
large items.
Each configuration $\sI \in \sF$ is associated with an integer $f(\sI)$ which
indicates that all large items from type $1$ to $f(\sI)$ were tested to be part
of $\sI$.
For example, if $\sI = (1,4,4,5)$ and $f(\sI) = 14$, then we know that a
configuration with one large item of type 1, two large items of type 4, and one
large item of type 5 is feasible, and that any large item of type up to 14 makes
$\sI$ infeasible if added to it.
Note that it may be the case that large items of type greater than 14 do not
make $\sI$ infeasible (the first phase was simply not able to decide it).
In the second phase the idea is to add items of type greater than $f(\sI)$ in
the remaining space of the bin by using a criterion of space: if the item's
area/volume is at most the remaining area/volume of the bin, such item will be
considered.
We may create an infeasible configuration at this point, however this is not an
obstacle since our goal is to find a configuration with maximum total weight.
Specifically, in the second phase we have an integer programming which simulates
a knapsack problem, described next.
Let $\sI \in \sF$ and let $\sI'$ be the set we want to build in the second
phase.
Let $\OR$ be the occupation ratio of bins containing small items, let $\OR(i)$
be the occupation ratio of Lbins of type $i$, and let $k$ be an integer such
that $\OR(i) > \OR$ for all $i > k$.
Our program will maximize $\sum_{1 \leq i \leq k} w(i)x(i)$ $+$ $(V -
\sum_{1 \leq i \leq k} v(i)x(i))/\OR$, where $w(i)$ is the weight of
large item $i$, $v(i)$ is the area/volume of item $i$ (according to the
dimension), $x(i)$ is a binary variable which indicates if item $i$ is in $\sI'$
or not, and $V$ is the area/volume of the bin (according to the dimension).
This program is subject to $x(i)=1$ for all $i \in \sI$, $x(i)=0$ for all $i
\leq f(\sI)$ such that $i \notin \sI$, and $\sum_{1 \leq i \leq k} v(i)x(i) \leq
V$.

This two-phase program is used in the following sections during the analysis of
each algorithm.


\section{Online Circle Packing in Squares}
\label{sec:cis}

For the Online Circle Packing in Squares problem, optimal values of $\rho_i$ are
known for $i \leq 30$, and good lower bounds are known for all $i$ up to
$9996$~\cite{packomania}.
Since $\rho_{9996} < 0.005076143$ and $\rho_i < \rho_{9996}$ for $i < 9996$, we
chose $M = \CISM$, which makes $2/M \leq 0.005555556$.
This means that we can find $K$ such that $\rho_{K+1} < 2/M \leq \rho_K$, and so
we can classify all large items.
Recall that $I_i$ denotes a large circle of type $i$.

We start by showing the occupation ratio of a closed Sbin, which is a result
valid for both bounded space and unbounded space algorithms.
Parameter $C$ is chosen to be $5$ in order to maximize this ratio, as
Theorem~\ref{thm:cis:occupation_ratio_small_C} shows.

\begin{theorem}
\label{thm:cis:occupation_ratio_small}
    The occupation ratio of a closed Sbin of type $i$, for $M \leq i < CM$, is
    at least $\left(1 - \frac{1}{C^2-1} - \frac{1}{C^2}\right)$ $\left(1 -
    \frac{8.62C}{M} + \frac{18.48}{M^2}\right)$ $\frac{\pi}{\sqrt{12}}
    \frac{M^2}{(M+1)^2}$.
\end{theorem}
\begin{proof}
    Note that the area loss in a closed Sbin is due to
      (i) packing circles into hexagons,
     (ii) tessellating a q-bin$(i,p)$ with hexagons, and
    (iii) non-full q-bins$(i,p)$.

    Consider (i). 
    We pack one circle of type $(i,p)$ into an hexagon of side length
    $\frac{4}{C^pi\sqrt{3}}$.
    Thus, the circle has area at least $\pi \left(\frac{2}{(i+1)C^p}\right)^2$
    and the hexagon has area area $\left(\frac{3\sqrt{3}}{2}\right)
    \left(\frac{4}{C^pi\sqrt{3}}\right)^2$.
    This gives an occupation ratio of at least $\frac{\pi}{\sqrt{12}}
    \frac{i^2}{(i+1)^2}$, which is at least $\frac{\pi}{\sqrt{12}}
    \frac{M^2}{(M+1)^2}$ because $i > M$ and $x^2/(x+1)^2$ is increasing for $x
    \geq 0$.

    Consider (ii).
    When we tessellate a q-bin$(i,p)$, we lose a strip of height at most once
    the height of the hexagon at the top border plus a strip of width at most
    twice the side of the hexagon at the right border, as depicted in
    Figure~\ref{fig:cis:area_loss}.
    Also, note that these strips intersect, so the length of one is $1/C^{p+1}$
    but the length of the other is not.
    See Figure~\ref{fig:cells} for the measurements of the hexagon.
    Thus, the area loss is at most $\left(\frac{1}{C^{p+1}}\right)
    \left(\frac{8}{C^p i \sqrt{3}}\right)$ $+$ $\left(\frac{1}{C^{p+1}} -
    \frac{8}{C^p i \sqrt{3}}\right) \left(\frac{4}{C^p i}\right)$ $=$
    $\frac{1}{C^{2p+1}i} \left(\frac{8}{\sqrt{3}}+4\right)$ $-$
    $\frac{1}{C^{2p}i^2} \left(\frac{32}{\sqrt{3}}\right)$ $<$
    $\frac{8.62}{C^{2p+1}M}$ $-$ $\frac{18.48}{C^{2p+2}M^2}$.
    If such tessellated q-bin$(i,p)$ is full, then its occupation ratio is at
    least $\frac{\pi}{\sqrt{12}} \frac{M^2}{(M+1)^2}$ $\left(\frac{1}{C^{2p+2}}
    - \frac{8.62}{C^{2p+1}M} +
    \frac{18.48}{C^{2p+2}M^2}\right)/\left(\frac{1}{C^{2p+2}}\right)$ $=$
    $\frac{\pi}{\sqrt{12}} \frac{M^2}{(M+1)^2}$ $\left(1 - \frac{8.62C}{M} +
    \frac{18.48}{M^2}\right)$.

\begin{figure}[h]
    \centering
    \resizebox{0.4\textwidth}{!}{\input{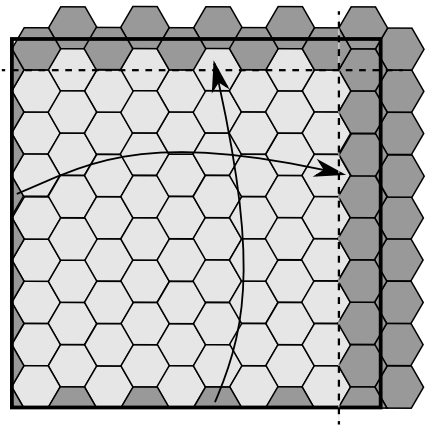}}
    \caption{For the Online Circle Packing in Squares, the darker hexagons are
    lost due to tiling one q-bin$(i,p)$ into hexagons. The borders of the
    q-bin$(i,p)$ are the thick lines. The area between the dashed lines and the
    border is the one being counted as lost. Note that it incorporates small
    pieces of hexagons lost in the left and bottom borders.}
    \label{fig:cis:area_loss}
\end{figure}

    Now, consider (iii).
    When an Sbin closes, there can be only one q-bin$(i,p)$ for each $p \geq 0$
    which is tessellated but not full.
    Also, no q-bin$(i,0)$ can be empty, but at most $C^2-1$ q-bins$(i,p)$ for
    each $p \geq 1$ can.
    Thus, the area loss due to non-full q-bins$(i,p)$ is at most $\sum_{p \geq
    0} (\frac{1}{C^{2p+2}})$ $+$ $\sum_{p \geq 1} (\frac{C^2-1}{C^{2p+2}}) =
    \frac{1}{C^2-1} + \frac{1}{C^2}$.

    Putting everything together, we have that the occupation ratio of the closed
    Sbin is at least $\left(1 - \frac{1}{C^2-1} - \frac{1}{C^2}\right)$ $\left(1
    - \frac{8.62C}{M} + \frac{18.48}{M^2}\right)$ $\frac{\pi}{\sqrt{12}}$
    $\frac{M^2}{(M+1)^2}$.
\end{proof}

\begin{theorem}
\label{thm:cis:occupation_ratio_small_C}
    The occupation ratio of a closed Sbin of type $i$, for $M \leq i < \CISC M$,
    is at least $\frac{\CISAA}{\CISAB}$ $\left(1 - \frac{\CISAC}{M} +
    \frac{\CISAD}{M^2}\right)$ $\frac{\pi}{\sqrt{12}}$ $\frac{M^2}{(M+1)^2}$.
\end{theorem}
\begin{proof}
    Given the result of Theorem~\ref{thm:cis:occupation_ratio_small} and using
    $M = \CISM$, we found that the maximum value of $\left(1 - \frac{1}{C^2-1} -
    \frac{1}{C^2}\right)$ $\left(1 - \frac{8.62C}{M} + \frac{18.48}{M^2}\right)$
    $\frac{\pi}{\sqrt{12}} \frac{M^2}{(M+1)^2}$, approximately \CISORsmall,
    happens when $C = \CISC$.
\end{proof}

Regarding large items, for the bounded space algorithm, note that each closed
Lbin of type $i$ packs $i$ circles of type $i$, which have radii at least
$\rho_{i+1}$.
Its occupation ratio is at least $i \pi \rho_{i+1}^2$.

Now we give the weighting function $w$ which will be used for the analysis of
the bounded space algorithm.
We define $w(I_i) = \frac{1}{i}$, so a closed Lbin of type $i$ (it has $i$
items) has total weight $\sum_{j=1}^i \frac{1}{i} = 1$.
Let $\OR$ $=$ $\frac{\CISAA}{\CISAB}$ $\left(1- \frac{\CISAC}{M} +
\frac{\CISAD}{M^1}\right)$ $\frac{\pi}{\sqrt{12}}$ $\frac{M^2}{(M+1)^2}$.
For a small circle $c$ of type $(i,p)$ and area $a(c)$, we define $w(c) =
\frac{a(c)}{\OR}$, so a closed Sbin $B$ of type $i$ has total weight $\sum_{c
\in B} \frac{a(c)}{\OR} = \frac{1}{\OR} a(B) \geq 1$, where $a(B)$ is the
sum of areas of items in $B$ and, according to
Theorem~\ref{thm:cis:occupation_ratio_small_C}, $a(B) \geq \OR$.

Theorem~\ref{thm:cis:bounded_ratio} concludes with the asymptotic competitive
ratio of the bounded space algorithm.

\begin{theorem}
\label{thm:cis:bounded_ratio}
    The algorithm for packing circles in squares with bounded space has
    asymptotic competitive ratio strictly below $\CISFactorBounded$.
\end{theorem}
\begin{proof}
    First note that, since $M = \CISM$, we have $\OR \geq \CISORsmall$. 
    We show some values of $i\pi\rho_{i+1}^2$ in
    Table~\ref{tab:cis:occupation_ratios}, from where we can note that for $i
    \geq 21$ we have $i\pi\rho_{i+1}^2 > \CISORsmall$.
    Thus, in the description of the two-phase program mentioned in the end of
    Section~\ref{sec:general}, the second phase has $k=20$, i.e., items of type
    up to 20 will be used to augment each configuration found during phase one.

\begin{table}
\renewcommand{\arraystretch}{1}
\setlength{\tabcolsep}{5pt}
\centering
\begin{tabular}{cc|cc|cc|cc}
$i$ & $i\pi\rho_{i+1}^2$ & $i$ & $i\pi\rho_{i+1}^2$ & $i$ & $i\pi\rho_{i+1}^2$ &
        $i$ & $i\pi\rho_{i+1}^2$ \\\hline
1  & 0.269506 & 38 & 0.790380 & 75 & 0.798059 & 112 & 0.819406 \\
2  & 0.406430 & 39 & 0.768280 & 76 & 0.799114 & 113 & 0.823647 \\
3  & 0.589049 & 40 & 0.773389 & 77 & 0.805433 & 114 & 0.822212 \\
4  & 0.539012 & 41 & 0.779668 & 78 & 0.810417 & 115 & 0.824644 \\
5  & 0.553297 & 42 & 0.768956 & 79 & 0.816878 & 116 & 0.827210 \\
6  & 0.573695 & 43 & 0.775801 & 80 & 0.812840 & 117 & 0.831850 \\
7  & 0.639593 & 44 & 0.771901 & 81  & 0.812682 & 118 & 0.837983 \\
8  & 0.698132 & 45 & 0.779857 & 82  & 0.811604 & 119 & 0.844561 \\
9  & 0.621032 & 46 & 0.772500 & 83  & 0.813538 & 120 & 0.828951 \\
10 & 0.637038 & 47 & 0.774662 & 84  & 0.818145 & 121 & 0.826427 \\
11 & 0.676929 & 48 & 0.775070 & 85  & 0.824701 & 122 & 0.823354 \\
12 & 0.676860 & 49 & 0.784267 & 86  & 0.808254 & 123 & 0.821206 \\
13 & 0.683131 & 50 & 0.792801 & 87  & 0.809045 & 124 & 0.823527 \\
14 & 0.711252 & 51 & 0.806713 & 88  & 0.804577 & 125 & 0.824775 \\
15 & 0.736311 & 52 & 0.799272 & 89  & 0.807785 & 126 & 0.825403 \\
16 & 0.690400 & 53 & 0.784604 & 90  & 0.809183 & 127 & 0.825023 \\
17 & 0.712728 & 54 & 0.785721 & 91  & 0.812688 & 128 & 0.821568 \\
18 & 0.712713 & 55 & 0.788024 & 92  & 0.809623 & 129 & 0.822238 \\
19 & 0.740519 & 56 & 0.789861 & 93  & 0.814375 & 130 & 0.822885 \\
20 & 0.717484 & 57 & 0.785549 & 94  & 0.811480 & 131 & 0.824018 \\
21 & 0.736604 & 58 & 0.789094 & 95  & 0.815584 & 132 & 0.825242 \\
22 & 0.730430 & 59 & 0.783854 & 96  & 0.819112 & 133 & 0.829802 \\
23 & 0.742673 & 60 & 0.788237 & 97  & 0.825048 & 134 & 0.831057 \\
24 & 0.753982 & 61 & 0.791144 & 98  & 0.831812 & 135 & 0.834326 \\
25 & 0.729297 & 62 & 0.798101 & 99  & 0.821731 & 136 & 0.838223 \\
26 & 0.743707 & 63 & 0.797034 & 100 & 0.819641 & 137 & 0.834469 \\
27 & 0.744288 & 64 & 0.803190 & 101 & 0.818015 & 138 & 0.837264 \\
28 & 0.752047 & 65 & 0.806944 & 102 & 0.818325 & 139 & 0.837932 \\
29 & 0.765618 & 66 & 0.812209 & 103 & 0.820286 & 140 & 0.838330 \\
30 & 0.752223 & 67 & 0.822742 & 104 & 0.824771 & 141 & 0.841156 \\
31 & 0.751754 & 68 & 0.804939 & 105 & 0.825170 & 142 & 0.845329 \\
32 & 0.764948 & 69 & 0.795970 & 106 & 0.816158 & 143 & 0.830805 \\
33 & 0.753806 & 70 & 0.794231 & 107 & 0.815139 & 144 & 0.831470 \\
34 & 0.758906 & 71 & 0.797673 & 108 & 0.816638 & 145 & 0.828671 \\
35 & 0.763582 & 72 & 0.797184 & 109 & 0.814799 & 146 & 0.826412 \\
36 & 0.762132 & 73 & 0.800550 & 110 & 0.817711 & 147 & 0.827501 \\
37 & 0.776068 & 74 & 0.795449 & 111 & 0.818423 & 148 & 0.830048 \\
\end{tabular}
\caption{For circles in square, these are the first 148 values of
        $i$ and the respective lower bound on the occupation ratio associated
        with $I_i$.}
\label{tab:cis:occupation_ratios}
\end{table}

    Now consider a bin of maximum weight.
    If it does not have any $I_1$ or $I_2$, phase two of the program finds a
    configuration with four $I_4$ and four $I_5$.
    Thus, the asymptotic competitive ratio is given by the sum of the weights of
    such circles ($4\frac{1}{4} + 4\frac{1}{5}$) plus the remaining area ($1 -
    4\pi\rho_5^2 - 4\pi\rho_6^2$) divided by $\OR$, and this gives a result
    which is less than $1.825165$.
    Now we need only to consider the configurations when these two types of
    circles are present.

    In the two-phase program, we tested all possible variations of
    configurations with $I_1$ and $I_2$, and we found the following ones:
    \begin{multicols}{3}
    \begin{enumerate}
        \item $I_1$, $I_2$, two $I_4$; \label{CIS:configA}
        \item $I_1$, $I_3$, two $I_4$; \label{CIS:configB}
        \item $I_1$, three $I_4$, $I_{12}$; \label{CIS:configC}
        \item two $I_2$, $I_3$, $I_5$, two $I_6$, $I_7$; \label{CIS:configD}
        \item two $I_2$, two $I_4$, nine $I_{20}$. \label{CIS:configE}
    \end{enumerate}
    \end{multicols}
    For example, during phase one we find that two $I_2$, one $I_3$, and one
    $I_5$ can be packed into a square.
    For such case, phase one is not able to say if adding an $I_6$ makes it
    infeasible or not, so phase two tries do add circles from type 6 to type 20
    on it, forming configuration~\ref{CIS:configD}.

    One can easily notice that removing items from the configurations above
    keeps them feasible (for instance a configuration with just one $I_1$), will
    give lower values of competitive ratios and this is why they are not
    considered in these calculations.

    In the following, we calculate the total sum of weights in each of these
    configurations, considering the formula $W + (1-V)/\OR$, as used by phase
    two of the program:
    \begin{enumerate}
       \item $1 + \frac{1}{2} + 2\frac{1}{4} +
                (1 - \pi\rho_2^2 - \pi\rho_3^2 - 2\pi\rho_5^2)/\OR$
                    $<$ $2.353507$;
        \item $1 + \frac{1}{3} + 2\frac{1}{4} +
                (1 - \pi\rho_2^2 - \pi\rho_4^2 - 2\pi\rho_5^2)/\OR$ 
                    $<$ $2.196255$;
        \item $1 + 3\frac{1}{4} + \frac{1}{12} +
                + (1 - \pi\rho_2^2 - 3\pi\rho_5^2 - \pi\rho_{13}^2)/\OR$
                    $<$ $2.203375$;
        \item $2\frac{1}{2} + \frac{1}{3} + \frac{1}{5} + 2\frac{1}{6} +
                \frac{1}{7} +
                (1 - 2\pi\rho_3^2 - \pi\rho_4^2 - \pi\rho_6^2 - 2\pi\rho_7^2
                    - \pi\rho_8^2)/\OR$ 
                    $<$ $2.014954$;
        \item $2\frac{1}{2} + 2\frac{1}{4} + 9\frac{1}{20} +
                (1 - 2\pi\rho_3^2 - 2\pi\rho_5^2 - 9\pi\rho_{21}^2)/\OR$ 
                    $<$ $1.951641$.
    \end{enumerate}
    Thus, the highest value we achieve is at most $2.353507$.
\end{proof}

\begin{figure}[h]
    \centering
    \resizebox{0.7\textwidth}{!}{\input{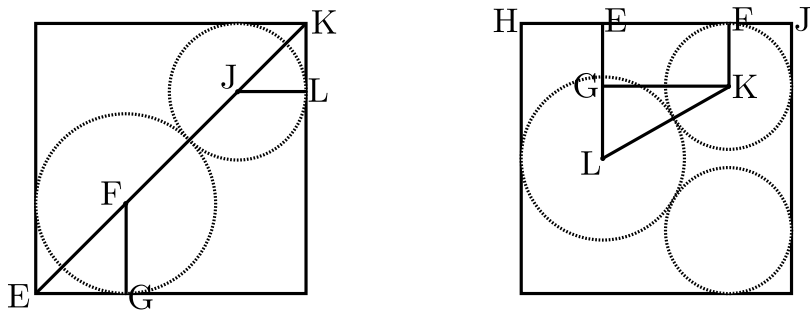}}
    \caption{For the Online Circle Packing in Squares, a waiting bin either places
    the $I_1$ at corner $E$ and the $I_2$ at corner $K$ (left drawing) or it
    places the $I_1$ centered to the left with one $I_4$ in corner $J$ and the
    other in the right bottom corner (right drawing).}
    \label{fig:cis_waiting_bin}
\end{figure}

Now we analyse the unbounded space algorithm.
We will use waiting bins that are supposed to pack either one $I_1$ with one
$I_2$ or one $I_1$ with two $I_4$, as depicted in
Figure~\ref{fig:cis_waiting_bin}.
All other circles are packed as in the bounded space algorithm.

Consider the notation of Figure~\ref{fig:cis_waiting_bin}.
In the left drawing, the circle centered at point $F$ is an $I_1$ while the
circle centered at point $J$ is an $I_2$.
We fix that the radius of the $I_1$ is $D$ and we want to find what is the
maximum radius $x$ that the $I_2$ can have as a function of $D$.
Note that, since $\angle EGF = 90\degree$ and $\angle GEF = 45\degree$, we have
$\overline{EF} = D\sqrt{2}$.
Similarly, $\overline{JK} = x\sqrt{2}$.
Since $\overline{EF} + \overline{VJ} + \overline{JK} = D\sqrt{2} + D+x +
x\sqrt{2} = \sqrt{2}$, we must have $x = \frac{\sqrt{2}}{\sqrt{2}+1}-D$.

For the right drawing of Figure~\ref{fig:cis_waiting_bin}, the circle centered
at point $L$ is an $I_1$ while the circle centered at point $K$ is an $I_4$.
Again, we fix that the radius of the $I_1$ is $D$ and we want to find what is
the maximum radius $x$ that the $I_4$ can have as a function of $D$.
Note that the triangle $LZK$ is rectangle, $\overline{GL} = \frac{1}{2} - x$
because $\overline{GE} = \overline{FK} = x$,~ $\overline{LK} = D+x$, and
$\overline{KG} = 1-(D+x)$ because $\overline{TE} = D$ and $\overline{FJ} = x$.
From $(D+x)^2 = (1-(D+x))^2 + (\frac{1}{2}-x)^2$ we have that $x = \frac{3}{2} -
\sqrt{2D+1}$.

Now note that we need to have $\rho_2 < D \leq \rho_1$, $\rho_3 <
\frac{\sqrt{2}}{\sqrt{2}+1}-D \leq \rho_2$, and $\rho_5 < \frac{3}{2} -
\sqrt{2D+1} \leq \rho_4$ so that the circles are of the desired types.
This is true for $\rho_2 < D < 0.331553$.
We fixed $D = \CISD$.

Finally, the algorithm works as follows.
If an $I_1$ with radius $r > D$ arrives, then we pack it as in the bounded space
algorithm (one per bin).
Otherwise, $r \leq D$ and we pack it in an already opened waiting bin containing
either one $I_2$ or at least one $I_4$, if they exist, or we open a new one,
pack the $I_1$ there, and let it open waiting for an $I_2$ or two $I_4$.
Note that this last case, in which we do not know if the next item that will
fill the waiting bin that contains one $I_1$ is one $I_2$ or one $I_4$, is the
reason why we allow the reorganization of items $I_1$ inside waiting bins.

Let $\gamma = \sqrt{2}/(\sqrt{2}+1) - D$.
If an $I_2$ with radius $r > \gamma$ arrives, then we pack it as in the bounded
space algorithm (two per bin).
When $r \leq \gamma$, the circles are labeled according to their arrival so that
the following steps can be repeated at every sequence of 72 of them. 
If the $I_2$ is among the first 70 of the sequence, then it is packed as in the
bounded space algorithm (two per bin);
if it is one of the last 2 of the sequence, then it is packed in a waiting bin:
either there is one with an $I_1$ or we open a new one, pack the circle there,
and let it open waiting for an $I_1$.

Let $\lambda = 3/2 - \sqrt{2D+1}$.
If an $I_4$ with radius $r > \lambda$ arrives, then we pack it as in the bounded
space algorithm (four per bin).
When $r \leq \lambda$, the circles are labeled according to their arrival so
that the following steps are repeated at every sequence of 34 such circles.
If the $I_4$ is among the first 32 of the sequence, then it is packed as in the
bounded space algorithm (four per bin);
if it is one of the last 2, then it is packed in a waiting bin.

We will use two weighting functions $w_1$ and $w_2$ for the analysis.
When the algorithm ends we have two possibilities: there are open waiting bins
with $I_1$, in which case we apply $w_1$ over all circles, or there are not, in
which case we apply $w_2$.
Both $w_1$ and $w_2$ differ from $w$ only regarding $I_1$, $I_2$, and $I_4$ if
their radii are at most $D$, $\gamma$, and $\lambda$, respectively.
If $I_1$ has radius $r \leq D$, then $w_1(I_1) = 1$ and $w_2(I_1) = 0$.
If $I_2$ has radius $r \leq \gamma$, then $w_1(I_2) =
\frac{\CISWoneCtwo}{\CISseqCtwo}$ and $w_2(I_2) =
\frac{\CISWtwoCtwo}{\CISseqCtwo}$.
If $I_4$ has radius $r \leq \lambda$, then $w_1(I_4) =
\frac{\CISWoneCfour}{\CISseqCfour}$ and $w_2(I_4) =
\frac{\CISWtwoCfour}{\CISseqCfour}$.

Note that, for both functions, the sum of weights in any bin is at least $1$ if
we do not consider $I_1$, $I_2$, and $I_4$, because this is true for $w$.
Now consider that $w_1$ was applied over such items. 
All waiting bins have total weight at least $1$ because they contain an $I_1$.
For every set of 72 $I_2$, we have 70 of them packed into 35 bins and the last 2
packed into 2 waiting bins with one $I_1$ each.
Thus, the average weight of these bins is $\left(35 \left(2\frac{35}{72}\right)
+ 2\left(1 + \frac{35}{72}\right)\right)/37 = 1$.
This is similar for $I_4$ and $w_2$.

Theorem~\ref{thm:cis:unbounded_ratio} concludes with the asymptotic competitive
ratio of our algorithm.

\begin{theorem}
\label{thm:cis:unbounded_ratio}
    The algorithm for packing circles in squares with unbounded space has
    asymptotic competitive ratio strictly below $\CISFactorUnbounded$.
\end{theorem}
\begin{proof}
    First consider $w_1$ was applied (this means there are open waiting bins
    with one $I_1$ each).
    An $I_1$ always has weight $1$ and we consider its area at least $\pi
    \rho_2^2$; an $I_2$ either has weight $\frac{1}{2}$ and area at least $\pi
    \gamma^2$ or it has weight $\frac{\CISWoneCtwo}{\CISseqCtwo}$ and area at
    least $\pi \rho_3^2$; and an $I_4$ either has weight $\frac{1}{4}$ and area
    at least $\pi \lambda^2$ or it has weight
    $\frac{\CISWoneCfour}{\CISseqCfour}$ and area at least $\pi \rho_5^2$.

    If $w_2$ was applied, an $I_1$ either has weight $1$ and area at least $\pi
    D^2$ or it has weight $0$ and area at least $\pi \rho_2^3$; an $I_2$ either
    has weight $\frac{1}{2}$ and area at least $\pi \gamma^2$ or it has weight
    $\frac{\CISWtwoCtwo}{\CISseqCtwo}$ and area at least $\pi \rho_3^2$; and an
    $I_4$ either has weight $\frac{1}{4}$ and area at least $\pi \lambda^2$ or
    it has weight $\frac{\CISWtwoCfour}{\CISseqCfour}$ and area at least $\pi
    \rho_5^2$.

    Now consider a bin of maximum weight.
    We again consider all configurations presented in the proof of
    Theorem~\ref{thm:cis:bounded_ratio} and recalculate the total sums of
    weights, but now with all possible combinations of weights and areas, as
    mentioned above.
    The last configuration is the one without circles of type 1 or type 2:
    \begin{enumerate}
       \item $1 + \frac{1}{2} + 2\frac{1}{4} +
                (1 - \pi\rho_2^2 - \pi\lambda^2 - 2\pi\gamma^2)/\OR$
                    $<$ $2.310294$,\\
             $1 + \frac{\CISWoneCtwo}{\CISseqCtwo} + 2\frac{1}{4} +
                (1 - \pi\rho_2^2 - \pi\rho_3^2 - 2\pi\gamma^2)/\OR$
                    $<$ $2.310033$,\\
             $1 + \frac{1}{2} + 2\frac{\CISWoneCfour}{\CISseqCfour} +
                (1 - \pi\rho_2^2 - \pi\lambda^2 - 2\pi\rho_5^2)/\OR$
                    $<$ $2.310467$,\\
             $1 + \frac{\CISWoneCtwo}{\CISseqCtwo} +
                    2\frac{\CISWoneCfour}{\CISseqCfour} +
                (1 - \pi\rho_2^2 - \pi\rho_3^2 - 2\pi\rho_5^2)/\OR$
                    $<$ $2.310206$,\\
             $1 + \frac{1}{2} + 2\frac{1}{4} +
                (1 - \pi D^2 - \pi\lambda^2 - 2\pi\gamma^2)/\OR$
                    $<$ $2.223957$,\\
             $1 + \frac{\CISWtwoCtwo}{\CISseqCtwo} + 2\frac{1}{4} +
                (1 - \pi D^2 - \pi\rho_3^2 - 2\pi\gamma^2)/\OR$
                    $<$ $2.251474$,\\
             $1 + \frac{1}{2} + 2\frac{\CISWtwoCfour}{\CISseqCfour} +
                (1 - \pi D^2 - \pi\lambda^2 - 2\pi\rho_5^2)/\OR$
                    $<$ $2.282954$,\\
             $1 + \frac{\CISWtwoCtwo}{\CISseqCtwo} +
                    2\frac{\CISWtwoCfour}{\CISseqCfour} +
                (1 - \pi D^2 - \pi\rho_3^2 - 2\pi\rho_5^2)/\OR$
                    $<$ $2.310470$;

        \item $1 + \frac{1}{3} + 2\frac{1}{4} +
                (1 - \pi\rho_2^2 - \pi\rho_4^2 - 2\pi\gamma^2)/\OR$ 
                    $<$ $2.166671$,\\
              $1 + \frac{1}{3} + 2\frac{\CISWoneCfour}{\CISseqCfour} +
                (1 - \pi\rho_2^2 - \pi\rho_4^2 - 2\pi\rho_5^2)/\OR$ 
                    $<$ $2.166843$,\\
              $1 + \frac{1}{3} + 2\frac{1}{4} +
                (1 - \pi D^2 - \pi\rho_4^2 - 2\pi\gamma^2)/\OR$ 
                    $<$ $2.080334$,\\
              $1 + \frac{1}{3} + 2\frac{\CISWtwoCfour}{\CISseqCfour} +
                (1 - \pi D^2 - \pi\rho_4^2 - 2\pi\rho_5^2)/\OR$ 
                    $<$ $2.139330$;

        \item $1 + 3\frac{1}{4} + \frac{1}{12} +
                (1 - \pi\rho_2^2 - 3\pi\gamma^2 - \pi\rho_{13}^2)/\OR$
                    $<$ $2.158998$,\\
              $1 + 3\frac{\CISWoneCfour}{\CISseqCfour} + \frac{1}{12} +
                (1 - \pi\rho_2^2 - 3\pi\rho_5^2 - \pi\rho_{13}^2)/\OR$
                    $<$ $2.159257$,\\
              $1 + 3\frac{1}{4} + \frac{1}{12} +
                (1 - \pi D^2 - 3\pi\gamma^2 - \pi\rho_{13}^2)/\OR$
                    $<$ $2.072661$,\\
              $1 + 3\frac{\CISWtwoCfour}{\CISseqCfour} + \frac{1}{12} +
                (1 - \pi D^2 - 3\pi\rho_5^2 - \pi\rho_{13}^2)/\OR$
                    $<$ $2.161156$;

        \item $2\frac{1}{2} + \frac{1}{3} + \frac{1}{5} + 2\frac{1}{6} +
                \frac{1}{7} +
                (1 - 2\pi\lambda^2 - \pi\rho_4^2 - \pi\rho_6^2 - 2\pi\rho_7^2
                    - \pi\rho_8^2)/\OR$ 
                    $<$ $1.987698$,\\
              $2\frac{\CISWoneCtwo}{\CISseqCtwo} + \frac{1}{3} + \frac{1}{5} +
                2\frac{1}{6} + \frac{1}{7} +
                (1 - 2\pi\rho_3^2 - \pi\rho_4^2 - \pi\rho_6^2 - 2\pi\rho_7^2
                    - \pi\rho_8^2)/\OR$ 
                    $<$ $1.987176$,\\
              $2\frac{\CISWtwoCtwo}{\CISseqCtwo} + \frac{1}{3} + \frac{1}{5} +
                2\frac{1}{6} + \frac{1}{7} +
                (1 - 2\pi\rho_3^2 - \pi\rho_4^2 - \pi\rho_6^2 - 2\pi\rho_7^2
                    - \pi\rho_8^2)/\OR$ 
                    $<$ $2.042731$;

        \item $2\frac{1}{2} + 2\frac{1}{4} + 9\frac{1}{20} +
                (1 - 2\pi\lambda^2 - 2\pi\gamma^2 - 9\pi\rho_{21}^2)/\OR$ 
                    $<$ $1.894801$,\\
              $2\frac{\CISWoneCtwo}{\CISseqCtwo} + 2\frac{1}{4} + 9\frac{1}{20} +
                (1 - 2\pi\rho_3^2 - 2\pi\gamma^2 - 9\pi\rho_{21}^2)/\OR$ 
                    $<$ $1.894279$,\\
              $2\frac{1}{2} + 2\frac{\CISWoneCfour}{\CISseqCfour} + 9\frac{1}{20} +
                (1 - 2\pi\lambda^2 - 2\pi\rho_5^2 - 9\pi\rho_{21}^2)/\OR$ 
                    $<$ $1.894974$,\\
              $2\frac{\CISWoneCtwo}{\CISseqCtwo} +
                2\frac{\CISWoneCfour}{\CISseqCfour} + 9\frac{1}{20} +
                (1 - 2\pi\rho_3^2 - 2\pi\rho_5^2 - 9\pi\rho_{21}^2)/\OR$ 
                    $<$ $1.894452$,\\
              $2\frac{\CISWtwoCtwo}{\CISseqCtwo} + 2\frac{1}{4} + 9\frac{1}{20} +
                (1 - 2\pi\rho_3^2 - 2\pi\gamma^2 - 9\pi\rho_{21}^2)/\OR$ 
                    $<$ $1.949835$,\\
              $2\frac{1}{2} + 2\frac{\CISWtwoCfour}{\CISseqCfour} + 9\frac{1}{20} +
                (1 - 2\pi\lambda^2 - 2\pi\rho_5^2 - 9\pi\rho_{21}^2)/\OR$ 
                    $<$ $1.953797$,\\
              $2\frac{\CISWtwoCtwo}{\CISseqCtwo} +
                2\frac{\CISWtwoCfour}{\CISseqCfour} + 9\frac{1}{20} +
                (1 - 2\pi\rho_3^2 - 2\pi\rho_5^2 - 9\pi\rho_{21}^2)/\OR$ 
                    $<$ $2.008831$;

        \item $4\frac{1}{4} + 4\frac{1}{5} +
                (1 - 4\pi\gamma^2 - 4\pi\rho_6^2)/\OR$
                $<$ $1.765996$,\\
              $4\frac{\CISWoneCfour}{\CISseqCfour} + 4\frac{1}{5} +
                (1 - 4\pi\rho_5^2 - 4\pi\rho_6^2)/\OR$
                $<$ $1.766342$,\\
              $4\frac{\CISWtwoCfour}{\CISseqCfour} + 4\frac{1}{5} +
                (1 - 4\pi\rho_5^2 - 4\pi\rho_6^2)/\OR$
                $<$ $1.883989$.
    \end{enumerate}
    And the result follows.
\end{proof}

\section{Online Circle Packing in Isosceles Right Triangles}
\label{sec:ciirt}

For the Online Circle Packing in Isosceles Right Triangles problem, optimal
values and good lower bounds of $\rho_i$ are known for $i \leq
299$~\cite{packomania}.
Since $\rho_{299} < 0.0211636617$ and $\rho_i < \rho_{299}$ for $i < 299$, we
chose $M = \CIIRTM$, which makes $2/M \leq 0.021505376$.
Recall that $I_i$ denotes a large circle of type $i$.

We also start by showing the occupation ratio of a closed Sbin in
Theorem~\ref{thm:ciirt:occupation_ratio_small}, which is a result valid for both
bounded space and unbounded space algorithms.
Parameter $C$ is chosen as $\CIIRTC$ in order to maximize this ratio.

\begin{theorem}
\label{thm:ciirt:occupation_ratio_small}
    The occupation ratio of a closed Sbin of type $i$, for $M \leq i < \CIIRTC
    M$, is at least $\frac{\CIIRTAA}{\CIIRTAB}$ $\left(1 - \frac{\CIIRTAC}{M} +
    \frac{\CIIRTAD}{M^2}\right)$ $\frac{\pi}{\sqrt{12}}$ $\frac{M^2}{(M+1)^2}$.
\end{theorem}
\begin{proof}
    Note that the area loss in a closed Sbin is due to
      (i) packing circles into hexagons,
     (ii) tessellating a q-bin$(i,p)$ with hexagons, and
    (iii) non-full q-bins$(i,p)$.

    Consider (i).
    Packing one circle of type $(i,p)$ of area at least $\pi
    \left(\frac{2}{(i+1)\CIIRTC^p}\right)^2$ into an hexagon of area
    $\frac{3\sqrt{3}}{2} \left(\frac{4}{\CIIRTC^pi\sqrt{3}}\right)^2$ gives an
    occupation ratio of at least $\frac{\pi}{\sqrt{12}} \frac{i^2}{(i+1)^2}$,
    which is at least $\frac{\pi}{\sqrt{12}} \frac{M^2}{(M+1)^2}$.

    For (ii), consider Figure~\ref{fig:ciirt:area_loss}.
    Note that line $r$ is perpendicular to two sides of the hexagons.
    Moving from the point where $r$ intersects the diagonal border of the
    triangle inwards, we notice that we lose a strip of width at most
    $\frac{4}{\CIIRTC^pi}$ (see Figure~\ref{fig:cells} for the measurements of
    the hexagon), whose extreme is depicted by dashed line $t$.
    Now note that $s$ is perpendicular to the diagonal border of the bin as well
    as to line $t$.
    From this, we note that the distance $\overline{AB}$, from the intersection
    $A$ of $t$ and the leg to the extreme $B$ of the triangle is
    $\frac{3+\sqrt{3}}{\CIIRTC^pi\sqrt{3}}$.
    The area of the q-bin$(i,p)$ minus the area of the triangle of leg length
    $\frac{1}{\CIIRTC^{p+1}} - \frac{3+\sqrt{3}}{\CIIRTC^pi\sqrt{3}}$ is the
    area of such lost strip.
    Also, note that we lose a strip of height at most half the height of the
    hexagon at the bottom border plus a strip of width at most half the side of
    the hexagon at the left border.

\begin{figure}[h]
    \centering
    \resizebox{0.4\textwidth}{!}{\input{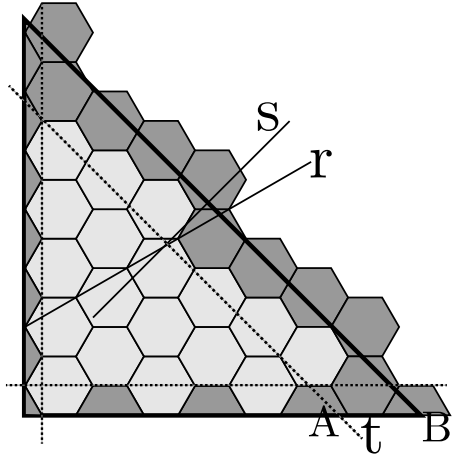}}
    \caption{For Online Circle Packing in Isosceles Right Triangle, the darker
    hexagons are lost due to tiling one q-bin$(i,p)$ into hexagons.  The borders
    of the q-bin$(i,p)$ are the thick lines. The area between the dashed lines
    and the borders is the one being counted as lost.}
    \label{fig:ciirt:area_loss}
\end{figure}

    Thus, the area loss is at most $\left(\frac{1}{2}\frac{1}{\CIIRTC^{2p+2}} -
    \frac{1}{2}(\frac{1}{\CIIRTC^{p+1}} -
    \frac{3+\sqrt{3}}{\CIIRTC^pi\sqrt{3}})^2\right)$ $+$ 
    $\left(\frac{1}{\CIIRTC^{p+1}} - \frac{3+\sqrt{3}}{\CIIRTC^pi\sqrt{3}}\right)
    \left(\frac{2}{\CIIRTC^pi}\right)$
    $+$
    $\left(\frac{1}{\CIIRTC^{p+1}} - \frac{3+\sqrt{3}}{\CIIRTC^pi\sqrt{3}}\right)
    \left(\frac{2}{\CIIRTC^pi\sqrt{3}}\right)$
    $=$
    $\frac{1}{\CIIRTC^{2p+1}i}\left(\frac{8}{\sqrt{3}}+4\right)$ $-$
    $\frac{1}{\CIIRTC^{2p}i^2}\left(16+\frac{28\sqrt{3}}{3}\right)$
    $<$
    $\frac{8.62}{\CIIRTC^{2p+1}M} - \frac{32.17}{\CIIRTC^{2p+2}M^2}$.
    If such tessellated q-bin$(i,p)$ is full, then its occupation ratio is at
    least $\frac{\pi}{\sqrt{12}} \frac{M^2}{(M+1)^2}$
    $(\frac{1}{2}\frac{1}{\CIIRTC^{2p+2}} - \frac{8.62}{\CIIRTC^{2p+1}M} +
    \frac{32.17}{\CIIRTC^{2p+2}M^2})/(\frac{1}{2}\frac{1}{\CIIRTC^{2p+2}})$
    $=$
    $\frac{\pi}{\sqrt{12}} \frac{M^2}{(M+1)^2}$ $\left(1 - \frac{25.86}{M} +
    \frac{32.17}{M^2}\right)$.

    Now, consider (iii).
    When an Sbin closes, there can be only one q-bin$(i,p)$ for each $p \geq 0$
    which is tessellated but not full.
    Also, no q-bin$(i,0)$ can be empty, but at most $\CIIRTC^2-1$ q-bins$(i,p)$
    for each $p \geq 1$ can.
    Thus, the area loss due to non-full q-bins$(i,p)$ is at most $\sum_{p \geq
    0} \left(\frac{1}{2}\frac{1}{\CIIRTC^{2p+2}}\right)$ $+$ $\sum_{p \geq 1}
    \left(\frac{1}{2}\frac{\CIIRTC^2-1}{\CIIRTC^{2p+2}}\right) =
    \frac{1}{2}\frac{1}{\CIIRTC^2-2} + \frac{1}{2}\frac{1}{\CIIRTC^2}$.

    Putting everything together, we have that the occupation ratio of the closed
    Sbin is at least $\left(\frac{1}{2} - \frac{1}{2}\frac{1}{\CIIRTC^2-1} -
    \frac{1}{2}\frac{1}{\CIIRTC^2}\right)$ $\left(1 - \frac{\CIIRTAC}{M} +
    \frac{\CIIRTAD}{M^2}\right)$ $\frac{\pi}{\sqrt{12}}$ $\frac{M^2}{(M+1)^2}$.
\end{proof}

Regarding the bounded space algorithm, note that each closed Lbin of type $i$
keeps $i$ circles of type $i$.
Thus, its occupation ratio is at least $i \pi \rho_{i+1}^2$.

Now we give the weighting function $w$ which will be used for the analysis of
the bounded space algorithm.
We define $w(I_i) = 1/i$, so a closed Lbin of type $i$ has total weight
exactly~$1$.
Let $\OR$ $=$ $\frac{\CIIRTAA}{\CIIRTAB}$ $\left(1- \frac{\CIIRTAC}{M} +
\frac{\CIIRTAD}{M^2}\right)$ $\frac{\pi}{\sqrt{12}}$ $\frac{M^2}{(M+1)^2}$.
For a small circle $c$ of type $(i,p)$ and area $a$, we define $w(c) =
a/\OR$, so a closed Sbin of type $i$ has total weight at least 1 since at
least $\OR$ of the bin's area is occupied, according to
Theorem~\ref{thm:ciirt:occupation_ratio_small}.

Theorem~\ref{thm:ciirt:bounded_ratio} concludes with the asymptotic competitive
ratio of the bounded space algorithm.

\begin{theorem}
\label{thm:ciirt:bounded_ratio}
    The algorithm for packing circles in isosceles right triangle with bounded
    space has asymptotic competitive ratio strictly below $\CIIRTFactorBounded$.
\end{theorem}
\begin{proof}
    Note that, since $M = \CIIRTM$, we have $\OR \geq \CIIRTORsmall$.
    We show some values of $i\pi\rho_{i+1}^2$ in
    Table~\ref{tab:ciirt:occupation_ratios}, from where we can note that for $i
    \geq 4$ we have $i\pi\rho_{i+1}^2 > \CIIRTORsmall$.
    Thus, in the two-phase program mentioned in the end of
    Section~\ref{sec:general}, the second phase can try to add items of type up to
    3 to augment each configuration found during phase one.
    However, phase one can easily verify if configurations of items of type up
    to 3 are feasible or not, so phase two is not necessary.

\begin{table}
\renewcommand{\arraystretch}{1}
\setlength{\tabcolsep}{5pt}
\centering
\begin{tabular}{cc|cc|cc|cc}
$i$ & $i\pi\rho_{i+1}^2$ & $i$ & $i\pi\rho_{i+1}^2$ & $i$ & $i\pi\rho_{i+1}^2$ &
        $i$ & $i\pi\rho_{i+1}^2$ \\\hline
1  & 0.134753 & 14 & 0.337587 & 27 & 0.360436 & 40 & 0.374184 \\
2  & 0.214343 & 15 & 0.333221 & 28 & 0.360441 & 41 & 0.373902 \\
3  & 0.241844 & 16 & 0.340169 & 29 & 0.362313 & 42 & 0.375378 \\
4  & 0.246066 & 17 & 0.339481 & 30 & 0.364711 & 43 & 0.375743 \\
5  & 0.285752 & 18 & 0.345218 & 31 & 0.365624 & 44 & 0.375958 \\
6  & 0.281614 & 19 & 0.348218 & 32 & 0.366048 & 45 & 0.378311 \\
7  & 0.291069 & 20 & 0.349180 & 33 & 0.367387 & 46 & 0.379287 \\
8  & 0.305438 & 21 & 0.350563 & 34 & 0.368083 & 47 & 0.378933 \\
9  & 0.319025 & 22 & 0.354141 & 35 & 0.369609 & 48 & 0.379238 \\
10 & 0.310476 & 23 & 0.353191 & 36 & 0.370060 & 49 & 0.381510 \\
11 & 0.318106 & 24 & 0.354329 & 37 & 0.371387 & 50 & 0.380398 \\
12 & 0.323289 & 25 & 0.358628 & 38 & 0.371382 & 51 & 0.382243 \\
13 & 0.329000 & 26 & 0.358762 & 39 & 0.374478 & 52 & 0.382699 \\
\end{tabular}
\caption{For circles in isosceles right triangle, these are the first 50 values
        of $i$ and the respective lower bound on the occupation ratio associated
        with $I_i$.}
\label{tab:ciirt:occupation_ratios}
\end{table}

    Now consider a bin of maximum weight.
    If it does not have any $I_1$ or $I_2$, phase two of the program finds a
    configuration with three $I_3$.
    Thus, the asymptotic competitive ratio is given by the sum of the weights of
    such circles ($3\frac{1}{3}$) plus the remaining area ($0.5 - 3\pi\rho_4^2$)
    divided by $\OR$, which is less than $2.049261$.
    Now we only need to consider the configurations when these two types of
    circles are present.

    With the first phase of the two-phase program, we tested all possible
    variations of configurations with items $I_1$, $I_2$, and $I_3$, and we found
    the following ones:
    \begin{multicols}{3}
    \begin{enumerate}
        \item $I_1$, $I_2$; \label{CIIRT:configA}
        \item $I_1$, two $I_3$; \label{CIIRT:configB}
        \item two $I_2$, $I_3$. \label{CIIRT:configC}
    \end{enumerate}
    \end{multicols}

    In the following, we calculate the total sum of weights in each of these
    configurations, considering the formula $W + (\frac{1}{2}-V)/\OR$:
    \begin{enumerate}
       \item $1 + \frac{1}{2} + (\frac{1}{2} - \pi\rho_2^2 - \pi\rho_3^2)/\OR$
                    $<$ $2.548932$;
        \item $1 + 2\frac{1}{3} + (\frac{1}{2} - \pi\rho_2^2 - 2\pi\rho_4^2)/\OR$ 
                    $<$ $2.495885$;
        \item $2\frac{1}{2} + \frac{1}{3} +
                (\frac{1}{2} - 2\pi\rho_3^2 - \pi\rho_4^2)/\OR$ 
                    $<$ $2.166716$.
    \end{enumerate}
    And so the result follows.
\end{proof}

\begin{figure}[h]
    \centering
    \resizebox{0.35\textwidth}{!}{\input{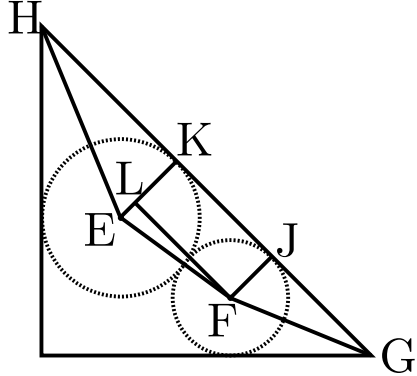}}
    \caption{For Online Circle Packing in Isosceles Right Triangle, a waiting
    bin places the $I_1$ at corner $E$ and the $I_2$ at corner $F$.}
    \label{fig:ciirt_waiting_bin}
\end{figure}

Now we analyse the unbounded space algorithm.
As circles of type greater than 3 already have an occupation ratio greater than
the occupation ratio of small circles, we will use waiting bins only to pack one
$I_1$ with one $I_2$, as depicted in Figure~\ref{fig:ciirt_waiting_bin}.
All other circles are packed as in the bounded space algorithm.

Consider the notation of Figure~\ref{fig:ciirt_waiting_bin}.
The circle centered at point $E$ is an $I_1$ while the circle centered at point
$F$ is an $I_2$.
We fix that the radius of the $I_1$ is $D$ and we want to find what is the
maximum radius $x$ that the $I_2$ can have as a function of $D$.
Note that, since $\angle KEH = 67.5\degree$, $\cos(67.5) =
\frac{\sqrt{2-\sqrt{2}} }{2}$, and triangle $KEH$ is rectangle, we have
$\overline{HK} = D\sqrt{\frac{2+\sqrt{2}}{2-\sqrt{2}} }$.
Similarly, $\overline{JG} = x\sqrt{\frac{2+\sqrt{2}}{2-\sqrt{2}} }$.
Since triangle $EFL$ is rectangle, $\overline{EL} = D-x$, $\overline{EF} = D+x$,
and $\overline{LF} = \overline{KJ}$, we have $\overline{KJ} = 2\sqrt{Dx}$.
At last, since $\overline{HK} + \overline{KJ} + \overline{JG} = \sqrt{2}$, we
must have $x = (5 - 4\sqrt{2})D - 2\sqrt{2(7 - 5\sqrt{2})D - 7\sqrt{2} +
10}\sqrt{D} - \sqrt{2} + 2$.

Now, for simplicity, let $\gamma$ $=$ $(5 - 4\sqrt{2})D$ $- 2\sqrt{2(7 -
5\sqrt{2})D - 7\sqrt{2} + 10}\sqrt{D}$ $- \sqrt{2} + 2$.
Note that we need to have $\rho_2 < D \leq \rho_1$ and $\rho_3 < \gamma \leq
\rho_2$ so that the circles are of the desired types.
This is true for $0.2071068 < D < 0.230249$ and so we fixed $D = \CIIRTD$.

The algorithm works as follows.
If an $I_1$ with radius $r > D$ arrives, then we pack it as in the bounded space
algorithm (one per bin).
Otherwise, $r \leq D$ and we pack it in an already opened waiting bin containing
one $I_2$, if it exists, or we open a new one, pack the $I_1$ there, and let it
open waiting for an $I_2$.

If an $I_2$ with radius $r > \gamma$ arrives, then we pack it as in the bounded
space algorithm (two per bin).
When $r \leq \gamma$, the circles are labeled according to their arrival so that
the following steps can be repeated at every sequence of 63 of them. 
The first 58 are packed in 29 bins (2 per bin) and the last five ones are packed
in five waiting bins when they arrive: either there is one with an $I_1$ or we
open one, pack the circle there, and let it open waiting for an $I_1$.

Again, we will use two weighting functions $w_1$ and $w_2$ for the analysis.
When the algorithm ends we also have two possibilities: there are open waiting
bins with $I_1$, in which case we apply $w_1$ over all circles, or there are
not, in which case we apply $w_2$.
Both $w_1$ and $w_2$ differ from $w$ only regarding $I_1$ and $I_2$, if
their radii are at most $D$ and $\gamma$, respectively.
If $I_1$ has radius $r \leq D$, $w_1(I_1) = 1$ and $w_2(I_1) = 0$.
If $I_2$ has radius $r \leq \gamma$, $w_1(I_2) =
\frac{\CIIRTWoneCtwo}{\CIIRTseqCtwo}$ and $w_2(I_2) =
\frac{\CIIRTWtwoCtwo}{\CIIRTseqCtwo}$.

It remains to show that, for both functions, the sum of weights in any bin is at
least $1$ on average.
Consider $w_1$ over items $I_1$ and $I_2$. 
All waiting bins have total weight at least $1$ because they contain an $I_1$.
For every set of 63 $I_2$, we have 58 of them packed into 29 bins and the last
five ones packed into five waiting bins with an $I_1$ each.
Thus, the average weight of these bins is
$\left(29(2\frac{\CIIRTWoneCtwo}{\CIIRTseqCtwo}) + 5(1 +
\frac{\CIIRTWoneCtwo}{\CIIRTseqCtwo})\right)/34 = 1$.
This is similar for $w_2$.

Theorem~\ref{thm:ciirt:unbounded_ratio} concludes with the asymptotic
competitive ratio of our algorithm.

\begin{theorem}
\label{thm:ciirt:unbounded_ratio}
    The algorithm for packing circles in isosceles right triangles with
    unbounded space has asymptotic competitive ratio strictly below
    $\CIIRTFactorUnbounded$.
\end{theorem}
\begin{proof}
    First consider $w_1$ was applied.
    An $I_1$ always has weight $1$ and we consider its area at least $\pi
    \rho_2^2$, and an $I_2$ either has weight $\frac{1}{2}$ and area at least $\pi
    \gamma^2$ or has weight $\frac{\CIIRTWoneCtwo}{\CIIRTseqCtwo}$ and area at
    least $\pi \rho_3^2$.

    If $w_2$ was applied, an $I_1$ either has weight $1$ and area at least
    $\pi D^2$ or has weight $0$ and area at least $\pi \rho_2^3$, and an $I_2$
    either has weight $\frac{1}{2}$ and area at least $\pi \gamma^2$ or has
    weight $\frac{\CIIRTWtwoCtwo}{\CIIRTseqCtwo}$ and area at least $\pi
    \rho_3^2$.

    Now consider a bin of maximum weight.
    We again consider all configurations presented in the proof of
    Theorem~\ref{thm:ciirt:bounded_ratio} and recalculate the total sums of
    weights, but now with all possible combinations of weights and areas, as
    mentioned above.
    The asymptotic approximation ratio when no $I_1$ or $I_2$ exist is still
    bounded above by $2.049261$.
    For the configurations with items $I_1$ or $I_2$ we now have:
    \begin{enumerate}
        \item $1 + \frac{1}{2} +
                (\frac{1}{2} - \pi\rho_2^2 - \pi\gamma^2)/\OR$
                    $<$ $2.509320$,\\
             $1 + \frac{\CIIRTWoneCtwo}{\CIIRTseqCtwo} +
                (\frac{1}{2} - \pi\rho_2^2 - \pi\rho_3^2)/\OR$
                    $<$ $2.509249$,\\
             $1 + \frac{1}{2} +
                (\frac{1}{2} - \pi D^2 - \pi\gamma^2)/\OR$
                    $<$ $2.430039$,\\
             $1 + \frac{\CIIRTWtwoCtwo}{\CIIRTseqCtwo} +
                (\frac{1}{2} - \pi D^2 - \pi\rho_3^2)/\OR$
                    $<$ $2.509334$.

        \item $1 + 2\frac{1}{3} +
                (\frac{1}{2} - \pi\rho_2^2 - 2\pi\rho_4^2)/\OR$ 
                    $<$ $2.495885$,\\
              $1 + 2\frac{1}{3} +
                (\frac{1}{2} - \pi D^2 - 2\pi\rho_4^2)/\OR$ 
                    $<$ $2.416605$;

        \item $2\frac{1}{2} + \frac{1}{3} +
                (\frac{1}{2} - 2\pi\gamma^2 - \pi\rho_4^2)/\OR$ 
                    $<$ $2.087492$,\\
              $2\frac{\CIIRTWoneCtwo}{\CIIRTseqCtwo} + \frac{1}{3} +
                (\frac{1}{2} - 2\pi\rho_3^2 - \pi\rho_4^2)/\OR$ 
                    $<$ $2.087351$,\\
              $2\frac{\CIIRTWtwoCtwo}{\CIIRTseqCtwo} + \frac{1}{3} +
                (\frac{1}{2} - 2\pi\rho_3^2 - \pi\rho_4^2)/\OR$ 
                    $<$ $2.246081$.
    \end{enumerate}
    And the result follows.
\end{proof}

\subsection{A lower bound for bounded space algorithms}
\label{sub:sec:ciirt:lower_bound}


Consider the following notation.
A circle $c$, of area $A(c)$, has weight $w(c) = 1/i$ if we can pack $i$ copies
of $c$ in a bin but we cannot pack $i+1$ copies of $c$.
For a set $\sC$ of circles, define $w(\sC) = \sum_{c \in \sC} w(c)$ and $A(\sC)
= \sum_{c \in \sC} A(c)$.
Hokama \textit{et al.}~\cite{2016-hokama-etal} showed that, for Online Circle
Packing in Squares, every bounded space online algorithm has competitive ratio
at least $w(\sC) + \frac{\sqrt{12}}{\pi} (1 - A(\sC))$, where $\sC$ is a set of
circles that can be packed into a unit square.
We can show the same result for Online Circle Packing in Isosceles Right
Triangles.

\begin{theorem}
\label{thm:ciirt:lower_bound_ratio}
    Let $\sC$ be a set of circles that can be packed into an isosceles right
    triangle of leg length one.
    Every bounded space online algorithm has competitive ratio at least $w(\sC)
    + \frac{\sqrt{12}}{\pi} (1 - A(\sC))$.
\end{theorem}

With the first phase of the program mentioned in the end of
Section~\ref{sec:general}, we verified that we can pack one circle of each type
1, 2, 4, 9, 11, and 28 in the same bin.
Using Theorem~\ref{thm:ciirt:lower_bound_ratio}, we have the following result.

\begin{theorem}
\label{thm:ciirt:lower_bound}
    Any bounded space online approximation algorithm for packing circles in
    isosceles right triangles has competitive ratio at least $\CIIRTLowerBound$.
\end{theorem}


\section{Online Sphere Packing in Cubes}
\label{sec:sic}

For Online Sphere Packing in Cubes, optimal values and good lower bounds of
$\rho_i$ are known for $i$ up to $1024192$~\cite{packomania}.
Since $\rho_{1024192} < 0.00554967$ and $\rho_i < \rho_{1024192}$ for $i <
1024192$, we chose $M = \SICM$, which makes $2/M \leq 0.006666667$.
Recall that $I_i$ denotes a large sphere of type $i$.

We start by showing the occupation ratio of a closed Sbin, which is a result
valid for both bounded space and unbounded space algorithms.
Parameter $C$ is chosen as $\SICC$ in order to maximize this ratio.

\begin{theorem}
\label{thm:sic:occupation_ratio_small}
    The occupation ratio of a closed Sbin of type $i$, for $M \leq i < \SICC M$,
    is at least $\frac{\SICAA}{\SICAB}$ $\left(1 - \frac{\SICAC}{M} +
    \frac{\SICAD}{M^2} - \frac{\SICAE}{M^3}\right)$ $\frac{\pi}{\sqrt{18}}$
    $\frac{M^3}{(M+1)^3}$.
\end{theorem}
\begin{proof}
    Note that the volume loss in a closed Sbin is due to three factors:
      (i) packing spheres into rhombic dodecahedra,
     (ii) tessellating a q-bin$(i,p)$ with rhombic dodecahedra, and
    (iii) non-full q-bins$(i,p)$.

    Consider (i).
    Packing one sphere of type $(i,p)$ of volume at least $\frac{4\pi}{3}
    \left(\frac{2}{\SICC^p(i+1)}\right)^3$ into a rhombic dodecahedron of volume
    $\frac{16\sqrt{3}}{9} \left(\frac{6}{\SICC^pi\sqrt{6}}\right)^3$ gives an
    occupation ratio of at least $\frac{\pi}{\sqrt{18}} \frac{i^3}{(i+1)^3}$
    which is at least $\frac{\pi}{\sqrt{18}} \frac{M^3}{(M+1)^3}$.

    For (ii), consider Figure~\ref{fig:sic:volume_loss}.
    When we tessellate a q-bin$(i,p)$, we lose three blocks of the bin.
    From face $GG'F'F$, a block of depth at most $\frac{4}{\SICC^pi}$, height
    $\frac{1}{\SICC^{p+1}}$, and width $\frac{1}{\SICC^{p+1}}$ is lost.
    From face $HEFG$, a block of depth at most $\frac{8}{\SICC^pi\sqrt{2}}$,
    height $\frac{1}{\SICC^{p+1}}$, and width $\frac{1}{\SICC^{p+1}} -
    \frac{4}{\SICC^pi}$ (because it intersects the last block) is lost.
    And from face $EFF'E'$, a block of depth at most $\frac{4}{\SICC^pi}$,
    height $\frac{1}{\SICC^{p+1}}-\frac{8}{\SICC^pi\sqrt{2}}$ (because it
    intersects the last block), and width
    $\frac{1}{\SICC^{p+1}}-\frac{4}{\SICC^pi}$ (because it intersects the first
    block) is lost.
    See Figure~\ref{fig:cells} for the measurements of the rhombic dodecahedron.

\begin{figure}[h]
    \centering
    \resizebox{\textwidth}{!}{\input{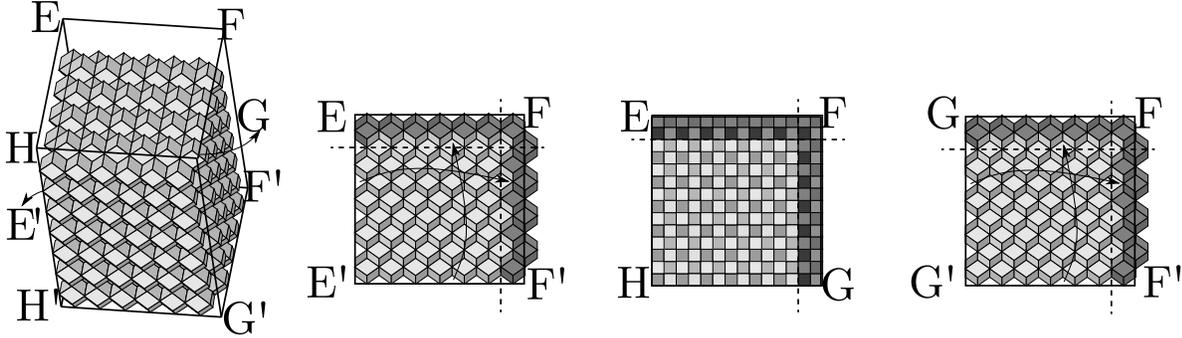}}
    \caption{For the Online Sphere Packing in Cubes, the darker rhombic
    dodecahedra are lost due to tessellating one q-bin$(i,p)$ into hexagons. The
    borders of the q-bin$(i,p)$ are the thick lines. The areas between the
    dashed lines and the borders are the ones being counted as lost. Note that
    they incorporate small pieces of rhombic dodecahedra in the other borders.}
    \label{fig:sic:volume_loss}
\end{figure}

    Thus, the volume loss due to the borders of one tessellated q-bin$(i,p)$ is
    at most
    $(\frac{4}{\SICC^pi})$ $(\frac{1}{\SICC^{p+1}})$ $(\frac{1}{\SICC^{p+1}})$
    $+$
    $(\frac{8}{\SICC^pi\sqrt{2}})(\frac{1}{\SICC^{p+1}})(\frac{1}{\SICC^{p+1}} -
    \frac{4}{\SICC^pi})$ $+$
    $(\frac{4}{\SICC^pi}) (\frac{1}{\SICC^{p+1}}-\frac{8}{\SICC^pi\sqrt{2}})
    (\frac{1}{\SICC^{p+1}}-\frac{4}{\SICC^pi})$ $=$
    $\frac{1}{\SICC^{3p+2}i}(4+\frac{8}{\sqrt{2}}+4)$ $-$
    $\frac{1}{\SICC^{3p+1}i^2}(\frac{32}{\sqrt{2}}+16+\frac{32}{\sqrt{2}})$ $+$
    $\frac{1}{\SICC^{3p}i^3}(\frac{128}{\sqrt{2}})$ $<$
    $\frac{13.66}{\SICC^{3p+2}M} - \frac{61.26}{\SICC^{3p+3}M^2} +
    \frac{90.51}{\SICC^{3p}M^3}$.
    If such tessellated q-bin$(i,p)$ is full, then its occupation ratio is at
    least 
    $\frac{\pi}{\sqrt{18}} \frac{M^3}{(M+1)^3}$
    $\left(\frac{1}{\SICC^{3p+3}} - \frac{13.66}{\SICC^{3p+2}M} +
    \frac{61.26}{\SICC^{3p+3}M^2} -
    \frac{90.51}{\SICC^{3p}M^3}\right)/\left(\frac{1}{\SICC^{3p+3}}\right)$
    $=$ $\frac{\pi}{\sqrt{18}} \frac{M^3}{(M+1)^3}$ $\left(1 - \frac{\SICAC}{M}
    + \frac{\SICAD}{M^2} - \frac{\SICAE}{M^3}\right)$.

    Now, consider (iii).
    When an Sbin closes, there can only be one q-bin$(i,p)$ for each $p \geq 0$
    which is tessellated but not full.
    Also, no q-bin$(i,0)$ can be empty, but at most $\SICC^3-1$ q-bins$(i,p)$
    for each $p \geq 1$ can.
    Thus, the volume loss due to non-full q-bins$(i,p)$ is at most $\sum_{p \geq
    0} (1/\SICC^{3p+3}) + \sum_{p \geq 1} ((\SICC^3-1)/\SICC^{3p+3}) =
    1/(\SICC^3-1) + 1/\SICC^3$.

    Putting everything together, we have that the occupation ratio of the closed
    Sbin of type $i$ is at least $\left(1 - \frac{1}{\SICC^3-1} -
    \frac{1}{\SICC^3}\right) \left(1 - \frac{\SICAC}{M} + \frac{\SICAD}{M^2} -
    \frac{\SICAE}{M^3}\right)$ $\frac{\pi}{\sqrt{18}}$ $\frac{M^3}{(M+1)^3}$.
\end{proof}

Regarding the bounded space algorithm, note that each closed Lbin of type $i$
keeps at most $i$ spheres of type $i$.
Thus, its occupation ratio is at least $i (4/3) \pi \rho_{i+1}^3$, since the
radius of any of its spheres is at least $\rho_{i+1}$.

The weighting function $w$ which will be used for the analysis is the same as
the ones given for the previous problems.
A large sphere $I_i$ receives weight $w(I_i) = 1/i$. 
For $\OR = \frac{\SICAA}{\SICAB} \left(1 - \frac{\SICAC}{M} +
\frac{\SICAD}{M^2} - \frac{\SICAE}{M^3}\right) \frac{\pi}{\sqrt{18}}
\frac{M^3}{(M+1)^3}$, a small sphere $s$ of type $(i,p)$ and radius $r$ receives
weight $w(s) = 4 \pi r^3 / (3\OR)$.

Theorem~\ref{thm:sic:bounded_ratio} concludes with the asymptotic competitive
ratio of the bounded space algorithm.

\begin{theorem}
\label{thm:sic:bounded_ratio}
    The algorithm for packing spheres in cubes with bounded space has asymptotic
    competitive ratio strictly below $\SICFactorBounded$.
\end{theorem}
\begin{proof}
    Since $M = \SICM$, we have $\OR \geq \SICORsmall$.
    We show some values of $i (4/3) \pi \rho_{i+1}^3$ in
    Table~\ref{tab:sic:occupation_ratios}, from where we can note that, for $i
    \geq 236$, we have $i(4/3)\pi\rho_{i+1}^3 > \SICORsmall$.
    We use the two-phase program mentioned in the end of
    Section~\ref{sec:general} to find possible configurations, which are shown
    below. 
    All configurations considered large spheres of type up to $235$ for the
    second phase.

\begin{table}
\renewcommand{\arraystretch}{1}
\setlength{\tabcolsep}{5pt}
\centering
\begin{tabular}{cc|cc|cc|cc}
$i$ & $i(4/3)\pi\rho_{i+1}^3$ & $i$ & $i(4/3)\pi\rho_{i+1}^3$ & $i$ &
        $i(4/3)\pi\rho_{i+1}^3$ & $i$ & $i(4/3)\pi\rho_{i+1}^3$ \\\hline
1  & 0.133417 & 38 & 0.512017 & 75  & 0.545362 & 211 & 0.592174 \\
2  & 0.210497 & 39 & 0.519548 & 76  & 0.537212 & 212 & 0.594981 \\
3  & 0.315745 & 40 & 0.509090 & 77  & 0.537329 & 213 & 0.597788 \\
4  & 0.308052 & 41 & 0.502683 & 78  & 0.537011 & 214 & 0.600594 \\
5  & 0.357008 & 42 & 0.500044 & 79  & 0.540779 & 215 & 0.603401 \\
6  & 0.393341 & 43 & 0.503414 & 80  & 0.538053 & 216 & 0.606207 \\
7  & 0.458148 & 44 & 0.512617 & 81  & 0.541092 & 217 & 0.609014 \\
8  & 0.418723 & 45 & 0.519972 & 82  & 0.540091 & 218 & 0.611820 \\
9  & 0.370946 & 46 & 0.528259 & 83  & 0.546666 & 219 & 0.614627 \\
10 & 0.374895 & 47 & 0.539743 & 84  & 0.553252 & 220 & 0.590387 \\
11 & 0.409321 & 48 & 0.510538 & 85  & 0.559839 & 221 & 0.588235 \\
12 & 0.446532 & 49 & 0.515781 & 86  & 0.566425 & 222 & 0.583545 \\
13 & 0.483743 & 50 & 0.515564 & 87  & 0.561817 & 223 & 0.584662 \\
14 & 0.417067 & 51 & 0.511943 & 88  & 0.542814 & 224 & 0.583463 \\
15 & 0.422827 & 52 & 0.515117 & 89  & 0.541957 & 225 & 0.584316 \\
16 & 0.450217 & 53 & 0.512968 & 90  & 0.541448 & 226 & 0.582069 \\
17 & 0.470755 & 54 & 0.511075 & 91  & 0.542754 & 227 & 0.582964 \\
18 & 0.463482 & 55 & 0.513913 & 92  & 0.546913 & 228 & 0.583569 \\
19 & 0.451938 & 56 & 0.522789 & 93  & 0.547168 & 229 & 0.585194 \\
20 & 0.466283 & 57 & 0.531881 & 94  & 0.549273 & 230 & 0.585598 \\
21 & 0.457616 & 58 & 0.541209 & 95  & 0.553838 & 231 & 0.587916 \\
22 & 0.467420 & 59 & 0.550540 & 96  & 0.558158 & 232 & 0.584947 \\
23 & 0.477859 & 60 & 0.559871 & 97  & 0.563516 & 233 & 0.584738 \\
24 & 0.475015 & 61 & 0.569203 & 98  & 0.567314 & 234 & 0.582294 \\
25 & 0.486897 & 62 & 0.578534 & 99  & 0.573024 & 235 & 0.584336 \\
26 & 0.504206 & 63 & 0.537287 & 100 & 0.561194 & 236 & 0.586733 \\
27 & 0.464907 & 64 & 0.538601 & 200 & 0.581381 & 237 & 0.589219 \\
28 & 0.482104 & 65 & 0.541275 & 201 & 0.579463 & 238 & 0.591670 \\
29 & 0.499323 & 66 & 0.533264 & 202 & 0.579480 & 239 & 0.594156 \\
30 & 0.516541 & 67 & 0.527819 & 203 & 0.580509 & 240 & 0.596642 \\
31 & 0.533759 & 68 & 0.531857 & 204 & 0.581582 & 241 & 0.599128 \\
32 & 0.495350 & 69 & 0.538335 & 205 & 0.578124 & 242 & 0.601614 \\
33 & 0.486560 & 70 & 0.546027 & 206 & 0.579691 & 243 & 0.604100 \\
34 & 0.496882 & 71 & 0.553801 & 207 & 0.581678 & 244 & 0.606586 \\
35 & 0.488672 & 72 & 0.544790 & 208 & 0.583755 & 245 & 0.609072 \\
36 & 0.499446 & 73 & 0.546273 & 209 & 0.586561 & 246 & 0.611558 \\
37 & 0.513083 & 74 & 0.553519 & 210 & 0.589368 & 247 & 0.614044 \\
\end{tabular}
\caption{For spheres in cubes, these are the values of $i$ and the
        respective lower bound on the occupation ratio associated with $i$ items
        $I_i$ for $1 \leq i \leq 100$ and $200 \leq i \leq 247$.}
\label{tab:sic:occupation_ratios}
\end{table}

    Consider a bin of maximum weight.
    If it does not have $I_1$ or $I_2$, then phase two of the program finds a
    configuration with three $I_3$, four $I_4$, five $I_5$, and one $I_{27}$.
    Thus, the asymptotic competitive ratio is given by the sum of the weights of
    such spheres ($3\frac{1}{3} + 4\frac{1}{4} + 5\frac{1}{5} + \frac{1}{27}$)
    plus the remaining volume ($1 - 4\pi\rho_4^3 - (16/3)\pi\rho_5^3 -
    (20/3)\pi\rho_6^3 - (4/3)\pi\rho_{28}^3$) divided by $\OR$, which is at
    most $3.040409$.

    All possible variations of configurations with $I_1$ and $I_2$ found by the
    two-phase program are the following:
    \begin{multicols}{3}
    \begin{enumerate}
        \item $I_1$, $I_2$, nine $I_9$, ten $I_{10}$, $I_{33}$;
                \label{SIC:configA}
        \item $I_1$, three $I_4$, one $I_8$, five $I_9$, ten $I_{10}$;
                \label{SIC:configB}
        \item two $I_2$, two $I_4$, five $I_5$, four $I_9$, three $I_{10}$.
                \label{SIC:configC}
    \end{enumerate}
    \end{multicols}

    In the following, we calculate the total sum of weights in each of these
    configurations, considering the formula $W + (1-V)/\OR$ and using $\OR
    \geq \SICORsmall$ as $\OR$ for all configurations:
    \begin{enumerate}
        \item $1 + \frac{1}{2} + 9\frac{1}{9} + 10\frac{1}{10} + \frac{1}{33} +
                (1 - (4/3)\pi\rho_2^3 - (4/3)\pi\rho_3^3 - (36/3)\pi\rho_{10}^3
                - (40/3)\pi\rho_{11}^3 - (4/3)\pi\rho_{34}^3)/\OR$
                    $<$ $3.531580$;
        \item $1 + 3\frac{1}{4} + \frac{1}{8} + 5\frac{1}{9} + 10\frac{1}{10} +
                (1 - (4/3)\pi\rho_2^3 - (12/3)\pi\rho_5^3 - (4/3)\pi\rho_9^3 -
                (20/3)\pi\rho_{10}^3 - (40/3)\pi\rho_{11}^3)/\OR$
                    $<$ $3.434356$;
        \item $2\frac{1}{2} + 2\frac{1}{4} + 5\frac{1}{5} + 4\frac{1}{9} +
                3\frac{1}{10} +
                (1 - (8/3)\pi\rho_3^3 - (8/3)\pi\rho_5^3 - (20/3)\pi\rho_6^3 -
                (16/3)\pi\rho_{10}^3 - (12/3)\pi\rho_{11}^3)/\OR$
                    $<$ $3.246382$.
    \end{enumerate}
\end{proof}

\begin{figure}[h]
    \centering
    \includegraphics[width=0.9\textwidth]{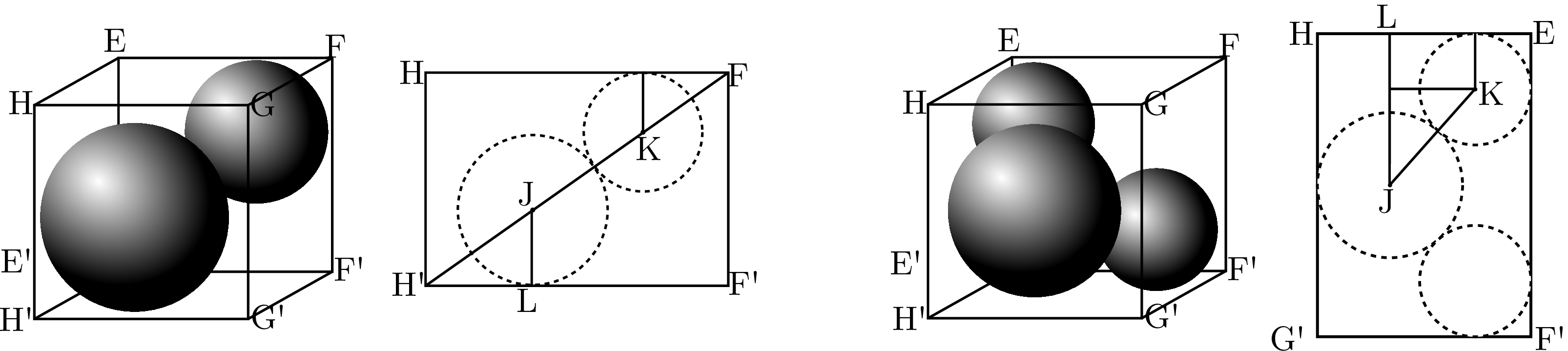}
    \caption{For Online Sphere Packing in Cubes, a waiting bin either places the
    $I_1$ at corner $H'$ and the $I_2$ at corner $F$ (left drawings) or it
    places the $I_1$ centered at face $HGG'H'$ with one $I_4$ at corner $E$ and
    the other at corner $F'$ (right drawings).}
    \label{fig:sic_waiting_bin}
\end{figure}

For the unbounded space algorithm, we also have that a waiting bin can pack one
$I_1$ with either one $I_2$ or two $I_4$, as depicted in
Figure~\ref{fig:sic_waiting_bin}.
All other spheres are packed as in the bounded space algorithm.

Consider the notation of Figure~\ref{fig:sic_waiting_bin}.
In the left drawings, the sphere centered at point $J$ is an $I_1$ while the
circle centered at point $K$ is an $I_2$.
We fix that the radius of the $I_1$ is $D$ and we want to find what is the
maximum radius $x$ that the $I_2$ can have as a function of $D$.
Because $\overline{HF} = \sqrt{2}$ and $\overline{HH'}$, we know $\overline{FH'}
= \sqrt{3}$.
So, $\sin(\angle JH'L) = \frac{1}{\sqrt{3}}$, which means $\overline{H'J} =
D\sqrt{3}$.
Similarly, $\overline{KF} = x\sqrt{3}$. 
Thus, we have $x = \frac{\sqrt{3}}{\sqrt{3}+1}-D$.

For the right drawings of Figure~\ref{fig:sic_waiting_bin}, the sphere centered
at point $J$ is an $I_1$ while the sphere centered at point $K$ is an $I_4$.
Because $\overline{HE} = 1$ and $\overline{TG'} = \sqrt{2}$ we can find that $x
= -\sqrt{2D + \sqrt{2}} + \frac{1}{\sqrt{2}} + 1$.

Let $\gamma = \frac{\sqrt{3}}{\sqrt{3}+1}-D$ and $\lambda = -\sqrt{2D + \sqrt{2}}
+ \frac{1}{\sqrt{2}} + 1$.
Again, we need to have $\rho_2 < D \leq \rho_1$, $\rho_3 < \gamma \leq \rho_2$,
and $\rho_5 < \lambda \rho_4$ so that the spheres are of the desired types.
This is true for $0.316987299 < D < 0.3342699108$.
We fixed $D = \SICD$.

The unbounded space algorithm works in the same way of the previous ones.
If an $I_1$ arrives and it has radius $r > D$, then we pack it as in the bounded
space algorithm (one per bin). 
Otherwise, if $r \leq D$, then we pack it in a waiting bin (favoring already
opened ones, as before).

If an $I_2$ with radius $r > \gamma$ arrives, then we pack it as in the bounded
space algorithm (two per bin).
Spheres of type $2$ with $r \leq \gamma$ are labeled according to their arrival
so that the following procedure can be repeated at every sequence of 29 of them. 
The first 28 spheres are packed in 14 bins (two per bin) and the last one is
packed in one waiting bin.

At last, if an $I_4$ with radius $r > \lambda$ arrives, then we pack it as in
the bounded space algorithm (four per bin).
Spheres of type $4$ with $r \leq \lambda$ are labeled according to their arrival
so that the following procedure can be repeated at every sequence of 18 such
spheres.
The first 16 are packed in 4 bins (4 per bin) and the last 2 is packed in a
waiting bin.

Again, when the algorithm finishes there can only be two possibilities: there
are open waiting bins with $I_1$ or not.
If the former happens, we apply function $w_1$ over all spheres, and if the
latter happens, we apply $w_2$.

Both $w_1$ and $w_2$ differ from function $w$ only regarding spheres of type
$1$, $2$, and $4$ of radii at most $D$, $\gamma$, and $\lambda$, respectively.
For $I_1$ of radius $r \leq D$, $w_1(s) = 1$ and $w_2(s) = 0$.
For $I_2$ of radius $r \leq \gamma$, $w_1(s) = \frac{\SICWoneStwo}{\SICseqStwo}$
and $w_2(s) = \frac{\SICWtwoStwo}{\SICseqStwo}$.
For $I_4$ of radius $r \leq \lambda$, $w_1(s) =
\frac{\SICWoneSfour}{\SICseqSfour}$ and $w_2(s) =
\frac{\SICWtwoSfour}{\SICseqSfour}$.

One can observe that, for both these functions, the sum of weights in any closed
bin is at least $1$ on average.
Theorem~\ref{thm:sic:unbounded_ratio} concludes with the asymptotic competitive
ratio of our algorithm.

\begin{theorem}
\label{thm:sic:unbounded_ratio}
    The algorithm for packing spheres in cubes with unbounded space has
    asymptotic competitive ratio strictly below $\SICFactorUnbounded$.
\end{theorem}
\begin{proof}
    If $w_1$ was applied, then an $I_1$ always has volume at least
    $(4/3)\pi\rho_2^3$ and it always has weight $1$ while if function $w_2$ was
    applied, then it has volume at least $(4/3)\pi D^3$ and also weight $1$.
    Regarding $I_2$, we can have either a sphere with weight
    $\frac{\SICWoneStwo}{\SICseqStwo}$ or $\frac{\SICWtwoStwo}{\SICseqStwo}$ and
    volume at least $(4/3)\pi\rho_3^3$ or spheres with weight $\frac{1}{2}$ and
    volume at least $(4/3)\pi\gamma^3$.
    And regarding $I_4$, we can have either spheres with weight
    $\frac{\SICWoneSfour}{\SICseqSfour}$ or $\frac{\SICWtwoSfour}{\SICseqSfour}$
    and volume at least $(4/3)\pi\rho_5^3$ or spheres with weight $\frac{1}{4}$
    and volume at least $(4/3)\pi\lambda^3$.

    Now consider a bin of maximum weight.
    We recalculate the total sums of weights, with all possible combinations of
    weights and volumes, as mentioned above, over the configurations presented
    in the proof of Theorem~\ref{thm:sic:bounded_ratio}.
    The last configuration is the one without spheres of type 1 or 2:
    \begin{enumerate}
        \item $1 + \frac{1}{2} + 9\frac{1}{9} + 10\frac{1}{10} + \frac{1}{33} +
                (1 - (4/3)\pi\rho_2^3 - (4/3)\pi\gamma^3 - (36/3)\pi\rho_{10}^3
                - (40/3)\pi\rho_{11}^3 - (4/3)\pi\rho_{34}^3)/\OR$
                    $<$ $3.514593$,\\
              $1 + \frac{\SICWoneStwo}{\SICseqStwo} + 9\frac{1}{9} +
                10\frac{1}{10} + \frac{1}{33} + 
                (1 - (4/3)\pi\rho_2^3 - (4/3)\pi\rho_3^3 - (36/3)\pi\rho_{10}^3
                - (40/3)\pi\rho_{11}^3 - (4/3)\pi\rho_{34}^3)/\OR$
                    $<$ $3.514339$,\\
              $1 + \frac{1}{2} + 9\frac{1}{9} + 10\frac{1}{10} + \frac{1}{33} + 
                (1 - (4/3)\pi D^3 - (4/3)\pi\gamma^3 - (36/3)\pi\rho_{10}^3 -
                (40/3)\pi\rho_{11}^3 - (4/3)\pi\rho_{34}^3)/\OR$
                    $<$ $3.480357$, or\\
              $1 + \frac{\SICWtwoStwo}{\SICseqStwo} + 9\frac{1}{9} +
                10\frac{1}{10} + \frac{1}{33} + 
                (1 - (4/3)\pi D^3 - (4/3)\pi\rho_3^3 - (36/3)\pi\rho_{10}^3 -
                (40/3)\pi\rho_{11}^3 - (4/3)\pi\rho_{34}^3)/\OR$
                    $<$ $3.514585$;

        \item $1 + 3\frac{1}{4} + \frac{1}{8} + 5\frac{1}{9} + 10\frac{1}{10} + 
                (1 - (4/3)\pi\rho_2^3 - (12/3)\pi\lambda^3 - (4/3)\pi\rho_9^3 -
                (20/3)\pi\rho_{10}^3 - (40/3)\pi\rho_{11}^3)/\OR$
                    $<$ $3.446663$,\\
              $1 + 3\frac{\SICWoneSfour}{\SICseqSfour} + \frac{1}{8} +
                5\frac{1}{9} + 10\frac{1}{10} + 
                (1 - (4/3)\pi\rho_2^3 - (12/3)\pi\rho_5^3 - (4/3)\pi\rho_9^3 -
                (20/3)\pi\rho_{10}^3 - (40/3)\pi\rho_{11}^3)/\OR$
                    $<$ $3.370020$,\\
              $1 + 3\frac{1}{4} + \frac{1}{8} + 5\frac{1}{9} + 10\frac{1}{10} + 
                (1 - (4/3)\pi D^3 - (12/3)\pi\lambda^3 - (4/3)\pi\rho_9^3 -
                (20/3)\pi\rho_{10}^3 - (40/3)\pi\rho_{11}^3)/\OR$
                    $3.412427$, or\\
              $1 + 3\frac{\SICWtwoSfour}{\SICseqSfour} + \frac{1}{8} +
                5\frac{1}{9} + 10\frac{1}{10} + 
                (1 - (4/3)\pi D^3 - (12/3)\pi\rho_5^3 - (4/3)\pi\rho_9^3 -
                (20/3)\pi\rho_{10}^3 - (40/3)\pi\rho_{11}^3)/\OR$
                    $<$ $3.502450$;

        \item $2\frac{1}{2} + 2\frac{1}{4} + 5\frac{1}{5} + 4\frac{1}{9} +
                3\frac{1}{10} +
                (1 - (8/3)\pi\gamma^3 - (8/3)\pi\lambda^3 - (20/3)\pi\rho_6^3 -
                (16/3)\pi\rho_{10}^3 - (12/3)\pi\rho_{11}^3)/\OR$
                    $<$ $3.207947$,\\
              $2\frac{\SICWoneStwo}{\SICseqStwo} + 2\frac{1}{4} + 5\frac{1}{5} +
                4\frac{1}{9} + 3\frac{1}{10} +
                (1 - (8/3)\pi\rho_3^3 - (8/3)\pi\lambda^3 - (20/3)\pi\rho_6^3 -
                (16/3)\pi\rho_{10}^3 - (12/3)\pi\rho_{11}^3)/\OR$ 
                    $<$ $3.207439$,\\
              $2\frac{1}{2} + 2\frac{\SICWoneSfour}{\SICseqSfour} + 5\frac{1}{5}
                + 4\frac{1}{9} + 3\frac{1}{10} +
                (1 - (8/3)\pi\gamma^3 - (8/3)\pi\rho_5^3 - (20/3)\pi\rho_6^3 -
                (16/3)\pi\rho_{10}^3 - (12/3)\pi\rho_{11}^3)/\OR$ 
                    $<$ $3.156852$,\\
              $2\frac{\SICWoneStwo}{\SICseqStwo} +
                2\frac{\SICWoneSfour}{\SICseqSfour} + 5\frac{1}{5} +
                4\frac{1}{9} + 3\frac{1}{10} +
                (1 - (8/3)\pi\rho_3^3 - (8/3)\pi\rho_5^3 - (20/3)\pi\rho_6^3 -
                (16/3)\pi\rho_{10}^3 - (12/3)\pi\rho_{11}^3)/\OR$ 
                    $<$ $3.156344$,\\
              $2\frac{\SICWtwoStwo}{\SICseqStwo} + 2\frac{1}{4} + 5\frac{1}{5} +
                4\frac{1}{9} + 3\frac{1}{10} + 
                (1 - (8/3)\pi\rho_3^3 - (8/3)\pi\lambda^3 - (20/3)\pi\rho_6^3 -
                (16/3)\pi\rho_{10}^3 - (12/3)\pi\rho_{11}^3)/\OR$ 
                    $<$ $3.276405$,\\
              $2\frac{1}{2} + 2\frac{\SICWtwoSfour}{\SICseqSfour} + 5\frac{1}{5}
                + 4\frac{1}{9} + 3\frac{1}{10} + 
                (1 - (8/3)\pi\gamma^3 - (8/3)\pi\rho_5^3 - (20/3)\pi\rho_6^3 -
                (16/3)\pi\rho_{10}^3 - (12/3)\pi\rho_{11}^3)/\OR$ 
                    $<$ $3.267963$, or\\
              $2\frac{\SICWtwoStwo}{\SICseqStwo} +
                2\frac{\SICWtwoSfour}{\SICseqSfour} + 5\frac{1}{5} +
                4\frac{1}{9} + 3\frac{1}{10} +
                (1 - (8/3)\pi\rho_3^3 - (8/3)\pi\rho_5^3 - (20/3)\pi\rho_6^3 -
                (16/3)\pi\rho_{10}^3 - (12/3)\pi\rho_{11}^3)/\OR$ 
                    $<$ $3.336421$;

        \item $3\frac{1}{3} + 4\frac{1}{4} + 5\frac{1}{5} + \frac{1}{27} +
                (1 - 4\pi\rho_4^3 - (16/3)\pi\lambda^3 - (20/3)\pi\rho_6^3 -
                (4/3)\pi\rho_{28}^3)/\OR$
                    $<$ $3.031488$,\\
              $3\frac{1}{3} + 4\frac{\SICWtwoSfour}{\SICseqSfour} + 5\frac{1}{5}
                + \frac{1}{27} +
                (1 - 4\pi\rho_4^3 - (16/3)\pi\rho_5^3 - (20/3)\pi\rho_6^3 -
                (4/3)\pi\rho_{28}^3)/\OR$
                    $<$ $2.929298$,\\
              $3\frac{1}{3} + 4\frac{\SICWtwoSfour}{\SICseqSfour} + 5\frac{1}{5}
                + \frac{1}{27} +
                (1 - 4\pi\rho_4^3 - (16/3)\pi\rho_5^3 - (20/3)\pi\rho_6^3 -
                (4/3)\pi\rho_{28}^3)/\OR$
                    $<$ $3.151520$.
    \end{enumerate}
    And the result follows.
\end{proof}

\subsection{A lower bound for bounded space algorithms}
\label{sub:sec:sic:lower_bound}

In this section we will consider the following notation.
A sphere $s$ has radius $r(s)$, weight $w(s) = 1/i$ if we can pack $i$ copies of
$s$ in a bin but we cannot pack $i+1$ copies of $s$, surface area $A(s)$, and
volume $V(s)$.
For a set $\sS$ of spheres, define $r(\sS) = \sum_{s \in \sS} r(s)$, $w(\sS) =
\sum_{s \in \sS} w(s)$, $A(\sS) = \sum_{s \in \sS} A(s)$, and $V(\sS) = \sum_{s
\in \sS} V(s)$.

\begin{theorem}
\label{thm:sic:lower_bound_ratio}
    Let $\sS$ be a set of spheres that can be packed into a unit cube.
    Every bounded space online algorithm has competitive ratio at least $w(\sS)
    + \frac{\sqrt{18}}{\pi} (1 - V(\sS))$.
\end{theorem}
\begin{proof}
    This proof is equal to the proof given by Hokama \textit{et
    al.}~\cite{2016-hokama-etal} in a similar result for packing circles in unit
    squares, with the proper alterations for spheres.

    Let $0 < \epsilon < 1$ be a constant, $\delta = \pi \epsilon /
    (4\sqrt{18})$, and $\varphi < \delta$.
    We will build a sequence of sets of spheres $\sS_0, \sS_1, \ldots$ until we
    obtain an $\sS_k$ such that $V(\sS_k) \geq (1 - \varphi)^2$.

    We start with $\sS_0 = \sS$ and, for each $n \geq 1$, the idea is to build
    $\sS_n$ by using the spheres in $\sS_{n-1}$ plus some small spheres of
    radius $\sqrt{6}\ell_n/3$ where $\ell_n < \ell_{n-1}$ and $\ell_1 < \min_{s
    \in \sS_0} r(s)$.
    We describe the details next.

    For $n \geq 1$, let $\ell_n \leq \varphi/(\frac{6}{3\sqrt{3}}A(\sS_{n-1}) +
    \frac{16\pi}{3}r(\sS_{n-1}) + \frac{32\pi}{9\sqrt{3}}|\sS_{n-1}| +
    \frac{12\sqrt{2}+44}{3\sqrt{3}})$.
    Suppose that $\sS_{n-1}$ can be packed in one bin and fix one of its
    packings.
    Consider a tessellation of rhombic dodecahedra of side $\ell_n$ on the bin
    in which $\sS_{n-1}$ is packed.
    Add to the packing spheres of radii $\sqrt{6}\ell_n/3$ in every feasible
    rhombic dodecahedron, i.e., rhombic dodecahedra that do not intersect
    spheres of $\sS_{n-1}$ or the borders of the bin.

    As in Theorem~\ref{thm:sic:occupation_ratio_small}, the total volume of
    infeasible rhombic dodecahedra that intersect the border of a unit cubic bin
    is at most $\frac{2\sqrt{2}\ell_n}{\sqrt{3}}$ $+$
    $\frac{4\ell_n}{\sqrt{3}}(1 - \frac{2\sqrt{2}\ell_n}{\sqrt{3}})$ $+$
    $\frac{2\sqrt{2}\ell_n}{\sqrt{3}} (1 - \frac{4\ell_n}{\sqrt{3}}) (1 -
    \frac{2\sqrt{2}\ell_n}{\sqrt{3}})$ $<$
    $(\frac{4\sqrt{2}+4}{\sqrt{3}})\ell_n$ $+$ $(\frac{32}{3\sqrt{3}})\ell_n^3$.
    If a rhombic dodecahedron intersects the interior of a sphere of radius $r$
    centered at a point $p$, then it is properly contained in the sphere of
    radius $r + 2\ell_n/\sqrt{3}$ centered at $p$ (see Figure~\ref{fig:cells}
    for the measurements of rhombic dodecahedra).
    Thus, the total volume of feasible rhombic dodecahedra is at least
{\footnotesize
  \setlength{\abovedisplayskip}{6pt}
  \setlength{\belowdisplayskip}{\abovedisplayskip}
  \setlength{\abovedisplayshortskip}{0pt}
  \setlength{\belowdisplayshortskip}{3pt}
    \begin{align}
            1 & - \left(\!\frac{4\sqrt{2}+4}{\sqrt{3}}\!\right)\!\ell_n -
            \left(\!\frac{32}{3\sqrt{3}}\!\right)\!\ell_n^3 -
            \!\!\sum_{s \in \sS_{n-1}} \left(\frac{4}{3} \pi \left(r(s) +
            \frac{2\ell_n}{\sqrt{3}}\right)^3\right) \nonumber \\
        &= 1 - \left(\!\frac{4\sqrt{2}+4}{\sqrt{3}}\!\right)\!\ell_n -
            \left(\!\frac{32}{3\sqrt{3}}\!\right)\!\ell_n^3 -
            \!\!\sum_{s \in \sS_{n-1}}\!\!\frac{4}{3} \pi \!\!\left(\!r(s)^3 +
            \frac{6}{\sqrt{3}}\ell_nr(s)^2 + 4\ell_n^2r(s) +
            \frac{8}{3\sqrt{3}}\ell_n^3\!\right) \nonumber \\
        &= 1 - \left(\!\frac{4\sqrt{2}+4}{\sqrt{3}}\!\right)\!\ell_n -
            \left(\!\frac{32}{3\sqrt{3}}\!\right)\!\ell_n^3 - V(\sS_{n-1}) -
            \frac{6}{3\sqrt{3}}\ell_n A(\sS_{n-1}) - \frac{16\pi}{3}\ell_n^2
            r(\sS_{n-1}) - \frac{32\pi}{9\sqrt{3}} \ell_n^3 |\sS_{n-1}|
            \nonumber \\
        &> 1 - \left(\!\frac{4\sqrt{2}+4}{\sqrt{3}}\!\right)\!\ell_n -
            \left(\!\frac{32}{3\sqrt{3}}\!\right)\!\ell_n - V(\sS_{n-1}) -
            \frac{6}{3\sqrt{3}}\ell_n A(\sS_{n-1}) - \frac{16\pi}{3}\ell_n
            r(\sS_{n-1}) - \frac{32\pi}{9\sqrt{3}} \ell_n |\sS_{n-1}|
            \label{sic:eq:ln_less_one} \\
        &\geq 1 - V(\sS_{n-1}) - \varphi \nonumber
    \end{align}
}%
    where~(\ref{sic:eq:ln_less_one}) follows because $\ell_n \leq 1$.

    Since every new sphere occupies a volume of $\frac{\pi}{\sqrt{18}}$ of the
    volume of the rhombic dodecahedron, we have that $V(\sS_n) \geq V(\sS_{n-1})
    + (1 - V(\sS_{n-1}) - \varphi)\frac{\pi}{\sqrt{18}}$.
    It follows, from induction, that $V(\sS_{n}) \geq
    (1-\varphi)\left(1-\left(1-\frac{\pi}{\sqrt{18}}\right)^n\right)$.
    Note that $\left(1-\frac{\pi}{\sqrt{18}}\right)^n < \varphi$ as $n$ grows,
    so there exists $k \geq 0$ such that $V(\sS_k) \geq (1-\varphi)^2$.
    From transitivity, $\sS_k$ can be packed in a bin and $\sS_0 \subseteq
    \sS_k$.

    Now consider an instance composed by $N$ disjoint copies of $\sS_k$ and let
    $\sS_0$ have $q_i$ spheres of type $i$, i.e., of radii between $\rho_{i+1}$
    and $\rho_i$.
    Note that an optimal offline solution uses $N$ bins to pack such instance
    and consider that, for the online algorithm, the spheres arrive in
    non-increasing order of radii.
    Any online algorithm with bounded space $B$ uses at least $Nq_i/i - B$ bins
    for every type of sphere in $\sS_0$. 
    Suppose that $n_j$ spheres, each of volume $v_j$, were added to $\sS_{j-1}$
    in order to construct $\sS_j$.
    Since the best packing of spheres has density $\pi/\sqrt{18}$, any online
    algorithm with bounded space $B$ uses at least $Nn_jv_j\sqrt{18}/\pi -
    B$ bins to pack the $n_j$ spheres. 
    
    Let $N \geq 2(t+k)B/\epsilon$, where $t$ is the number of different types of
    spheres in $\sS$.
    By the analysis above, any algorithm with bounded space $B$ uses at least
{\footnotesize
  \setlength{\abovedisplayskip}{6pt}
  \setlength{\belowdisplayskip}{\abovedisplayskip}
  \setlength{\abovedisplayshortskip}{0pt}
  \setlength{\belowdisplayshortskip}{3pt}
    \begin{align*}
        \sum_{\textrm{type}~i~\textrm{in}~\sS_0} &\left(\frac{Nq_i}{i} -
            B\right) + \sum_{j=1}^k \left(Nn_jv_j \frac{\sqrt{18}}{\pi} -
            B\right) 
        = N\sum_{\textrm{type}~i~\textrm{in}~\sS_0} \frac{q_i}{i} - tB +
            N\frac{\sqrt{18}}{\pi} \sum_{j=1}^k (n_jv_j) - kB \\
        &= N\!\!\left(\!\sum_{\textrm{type}~i~\textrm{in}~\sS_0}\! \frac{q_i}{i}
            \!+\! \frac{\sqrt{18}}{\pi} \!(V(\sS_k) - V(\sS_0))\!\!\right) \!-\!
            (t\!+\!k)B 
        \geq N\!\!\left(\!w(\sS) \!+\!
            \frac{\sqrt{18}}{\pi}\!\left((1-\varphi)^2 \!-\!
            V(\sS)\right)\!\!\right) \!-\! (t\!+\!k)B \\
        &> N\!\!\left(\!w(\sS) \!+\!  \frac{\sqrt{18}}{\pi}\!\left(1 \!-\!
            2\varphi \!-\!  V(\sS)\right)\!\!\right) \!-\! (t\!+\!k)B 
        = N\!\!\left(\!w(\sS) \!+\! \frac{\sqrt{18}}{\pi}\!(1 \!-\!
            V(\sS))\!\!\right) \!-\!  N\frac{\sqrt{18}}{\pi}2\varphi \!-\!
            (t\!+\!k)B \\
        &> N\!\!\left(\!w(\sS) \!+\! \frac{\sqrt{18}}{\pi}\!(1 \!-\!
            V(\sS))\!\!\right) \!-\!  \frac{N\epsilon}{2} \!-\! (t\!+\!k)B 
        \geq N\!\!\left(\!w(\sS) \!+\! \frac{\sqrt{18}}{\pi}\!(1 \!-\!
            V(\sS))\!\!\right) \!-\!  N\epsilon \enspace,
    \end{align*}
}%
    and so the result follows.
\end{proof}

With the first phase of the program mentioned in the end of
Section~\ref{sec:general}, we verified that we can pack one sphere of type 1,
one of type 2, four of type 9, and one of type 18 in the same bin.
Using Theorem~\ref{thm:sic:lower_bound_ratio}, we have the following result.

\begin{theorem}
\label{thm:sic:lower_bound}
    Any bounded space online approximation algorithm for packing spheres in
    cubes has competitive ratio at least $\SICLowerBound$.
\end{theorem}

\bibliographystyle{plain}
\bibliography{refs}

\end{document}